\documentclass[journal]{new-aiaa} 
\usepackage[utf8]{inputenc}

\usepackage{graphicx}
\usepackage{epsfig,dcolumn,amsmath,amssymb}
\usepackage[version=4]{mhchem}
\usepackage{siunitx}
\usepackage{longtable,tabularx}
\usepackage{subfigure}
\usepackage{subfigmat}
\usepackage[table]{xcolor}
\usepackage{multirow}
\usepackage{hhline}
\usepackage{booktabs}
\usepackage{caption}
\usepackage{arydshln}
\usepackage{upgreek}
\usepackage[title]{appendix}
\def\xstrut{\rule{0pt}{2ex}}    

\def\bs #1{\ensuremath{{\boldsymbol{#1}}}}
\DeclareMathOperator*{\argmin}{arg\,min}
\def\xstrut{\rule{0pt}{1.5ex}}
\def\dcm(#1,#2,#3){\ensuremath{{\bf {#1}}_{#2}^{#3}}}

\def\norm(#1){\ensuremath{\lVert {#1} \rVert}}
\def\wnorm(#1,#2){\ensuremath{{\lVert {#1} \rVert}_{\xstrut {#2}}}}
\def\w2norm(#1,#2){\ensuremath{{\left\Vert {#1} \right\Vert}^2_{\xstrut {#2}}}}
\def\deg(#1){\ensuremath{{#1}^{\circ}}}

\title{A Two-Stage Batch Algorithm for \\ Nonlinear Static Parameter Estimation}

\author{Kerry Sun\footnote{Ph.D. Candidate, Department of Aerospace Engineering and Mechanics; sunx0486@umn.edu, Student Member AIAA.} and Demoz Gebre-Egziabher \footnote{Professor, Department of Aerospace Engineering and Mechanics; gebre@umn.edu, Associate Fellow AIAA.}}
\affil{University of Minnesota, Twin Cities, Minneapolis, Minnesota, 55455}

\begin{document}

\maketitle

\begin{abstract}
A two-stage batch estimation algorithm for solving a class of nonlinear, static parameter estimation problems that appear in aerospace engineering applications is proposed. It is shown how these problems can be recast into a form suitable for the proposed two-stage estimation process. In the first stage, linear least squares is used to obtain a subset of the unknown parameters (set 1), while a residual sampling procedure is used for selecting initial values for the rest of the parameters (set 2). In the second stage, depending on the uniqueness of the local minimum, either only the parameters in the second set need to be re-estimated, or all the parameters will have to be re-estimated simultaneously, by a nonlinear constrained optimization. The estimates from the first stage are used as initial conditions for the second stage optimizer. It is shown that this approach alleviates the sensitivity to initial conditions and minimizes the likelihood of converging to an incorrect local minimum of the nonlinear cost function. An error bound analysis is presented to show that the first stage can be solved in such a way that the total cost function will be driven to the optimal cost, and the difference has an upper bound.  Two tutorial examples are used to show how to implement this estimator and compare its performance to other similar nonlinear estimators.  Finally, the estimator is used on a 5-hole Pitot tube calibration problem using flight test data collected from a small Unmanned Aerial Vehicle (UAV) which cannot be easily solved with single-stage methods. 
\end{abstract}

\section*{Nomenclature}

{\renewcommand\arraystretch{1.0}
\noindent\begin{longtable*}{@{}l @{\quad=\quad} l@{}} 
$a_x, a_y, a_z$ & body-axis translational acceleration \\
$b_{a_x}, b_{a_y}, b_{a_z}$ & bias of body-axis translational acceleration \\
$b_{p}, b_{q}, b_{r}$ & bias of body-axis rotational velocity \\
$g$ & gravitational acceleration \\
${\bf f}$  & nonlinear dynamic model \\
${\bf h}$  & nonlinear measurement model \\
$K_{\alpha}, K_{\alpha}$ & sensitivity coefficients of angle-of-attack and sideslip \\ 
$p,q,r$    & body-axis rotational velocity \\
$P_s, P_t$ & static and dynamic pressures \\  
$P_{\Delta \alpha}$ & differential angle-of-attack pressure \\
$P_{\Delta \beta}$ &  differential sideslip angle pressure \\
${\bf R}$ &  noise covariance matrix\\
$t$        & time \\
$u,v,w$    & body-axis translational velocity \\
${\bf u}$  & input vector \\
$V_a$      & airspeed \\ 
${\bf x}$  & state vector \\
${\bf y}$  & true output vector \\
$\alpha$   & angle-of-attack \\
$\beta$    & sideslip angle \\
$\rho$     & air density \\ 
$\phi, \theta, \psi$ & Euler angles \\ 
$N\left(\mu,\sigma^2\right)$ & Normal (Gaussian) distribution with mean $\mu$ and standard deviation $\sigma$ \\
\end{longtable*}}
	    \par\noindent\textit{\small{Superscripts}} 
{\renewcommand\arraystretch{1.0}
	\noindent\begin{longtable*}{@{}l @{\quad=\quad} l@{}}	    
		$(\cdot)^T$ & transpose  \\
		$(\cdot)^{-1}$ & matrix inverse  \\
		$\hat{(\cdot)}$ & estimate of $(\cdot)$  \\
		$(\cdot)^*$ & optimal value of $(\cdot)$  \\
\end{longtable*}}

\section{Introduction}

\lettrine{T}his paper presents an algorithm for solving a class of nonlinear estimation problems that appear in aerospace guidance, navigation and control. These nonlinear estimation problems appear in applications such as vehicle system identification; sensor calibration; and vehicle positioning, navigation and timing (PNT). In the past, these problems have been solved either by standard estimators (e.g., the Kalman filter or its many variants \cite{Stengel,Simon2006}; maximum likelihood estimators \cite{Kay1993}; or output-error minimization  \cite{Jategaonkar2006,Klein2006,Grauer2015}) or, in many instances, by \emph{ad hoc} approaches developed for the particular problem at hand. It is the claim of this paper that a large number of these nonlinear estimation problems have a similar mathematical structure which can be exploited in a two-stage estimator. This estimator can overcome the initial condition sensitivity problem, have good convergence, and, in many instances, have a guaranteed estimation bound on the total cost function. In this paper, we describe this nonlinear mathematical structure and discuss why it arises in many aerospace sensing and estimation problems.  Subsequently, we develop an estimator designed to exploit this nonlinear structure and provide examples to demonstrate its performance.

The class of nonlinear estimation problems that are the subject of this paper have the following form:
\begin{align}
	{\bf z}_k = {\bf A}\left({\bs \xi}_2\right){\bs \xi}_1 + {\bf b}\left({\bs \xi}_2\right) + {\bf v}_k
	\label{eq:canonical_structure}
\end{align}
where ${\bs \xi} = {\begin{bmatrix} {\bs \xi}_1^T & {\bs \xi}_2^T\end{bmatrix}}^T$ is the vector of parameters to be estimated, ${\bf z}_k$ is a measurement vector at any discrete time $t_k$ and ${\bf v}_k$ is the noise vector corrupting the measurement at $t_k$.  The matrix ${\bf A}$ and the vector ${\bf b}$ are functions of the unknown parameters ${\bs \xi}_2$ only. This mathematical form appears often in parameter estimation problems.  As we show later in the paper, this form arises when embedded in the problem at hand is the standard sensor error model which relates \emph{measured} quantities ${\bf z}_k$ to their \emph{true} values ${\bf y}_{k}$ given by the following mathematical relationship from Ref. \cite[Eq. (10.13)]{Klein2006} and 
\cite[Eq (4.15), (4.16) and (4.17)]{Groves}
\begin{align}
	{\bf z}_k =  {\bf h}_k\left({\bf y}_k,\boldsymbol{\xi}\right) + {\bf v}_k =  {\bf C}\,{\bf y}_{k} + {\bf n}_k + {\bf v}_k
	\label{eq:error_model}
\end{align}
In the standard sensor error model given above, the matrix ${\bf C}$ is a matrix whose entries are a function of unknown sensor parameters (e.g., scale factor errors, axis misalignment errors, etc., ), the vector ${\bf n}_k$ consists of unknown null-shifts (biases). Both ${\bf C}$ and  ${\bf n}_k$ are functions of the parameter $\boldsymbol{\xi}$. The vector ${\bf v}_k$ is independent Gaussian white measurement noise.  In the appendices of this paper, we provide a general canonical form and two examples that show how the form of Eq. (\ref{eq:canonical_structure}) arises from Eq. (\ref{eq:error_model}).  

The algorithm proposed in this paper exploits the structure in Eq. (\ref{eq:canonical_structure}) by using a two-stage estimation scheme.  In the first stage, we solve a linear least squares problem for the parameter vector ${\bs \xi}_1$, where the remainder of the unknowns in the parameter vector ${\bs \xi}_2$ are held fixed at some pre-determined values (i.e. using prior knowledge or systematically selected).  In the second stage, depending on the uniqueness of the local minimum, we solve a constrained nonlinear optimization problem for either ${\bs \xi}_2$ only (and ${\bs \xi}_1 $ can be determined consequently), or all of the unknowns (${\bs \xi}_1$ and ${\bs \xi}_2$) simultaneously, by using the estimates from the first-stage as the initial conditions for the optimization.  As will be demonstrated later, this formulation overcomes initial condition sensitivity issues and leads to excellent convergence properties and, in many instances, guaranteed upper bounds on the cost function.

\subsection{Prior Work}

The idea of solving nonlinear estimation problems in two stages is not new and some of the earliest work relevant to the discussion here dates from the early 1970's \cite{Lawton1971,Guttman1973,Golub1973}. In particular, Golub and Pereyra \cite{Golub1973} dealt with a nonlinear parameter estimation problem by solving only a subset of the total parameters in the first stage. They used the idea of removing ``conditionally linear" parameters to separate linear and nonlinear parameters \cite{Bates1988}. It was proved that all the critical points (local or global optima) of the first stage yield the same critical points as the nonlinear least squares problem. When the nonlinear estimate is solved in the first stage, then the rest of the unknown can be solved for linearly. However, the numerical algorithm can be complex as it requires computing special derivatives of orthogonal projectors that have to be obtained for the efficient gradient descent optimization method to work. 

Haupt and Kasdin \cite{Haupt1996} proposed a two-step, recursive and iterative estimation algorithm. The algorithm uses a change of variables to split the cost function into a linear problem in the first step and a nonlinear problem in the second step. The split is done in such a way that the first-step states become measurements for the second-step states.  While this estimator is powerful and has been used successfully in many aerospace estimation problems, the underlying approach will not always lead to an optimal estimate, most notably when the second step cost function is non-convex.  Furthermore, as we show later, it is not always obvious (or even possible) how to split some problems into a linear and nonlinear step by a simple change of variables.

 Another similar and highly effective two-step procedure was proposed by Alonso and Shuster \cite{Alonso2002} to solve the magnetometer calibration problem. Their approach ``centers'' the nonlinear measurement model into a linear model, and solves a centered estimate in the first step. In the second step, it uses the centered estimate as an initial estimate to approximate the original estimated parameters. However, this algorithm is somewhat \emph{ad hoc} in that it is very specific to the magnetometer calibration problem; the statistical properties of the estimation errors cannot be easily transferred to other general estimation problems. The Prony algorithm \cite{Hauer1990} is another example of an \emph{ad hoc} estimation approach that has been used successfully in the problem of estimating frequency, amplitude, phase and damping components of electrical power system response signals. 

In the field of aerodynamic parameter estimation, the equation-error approach \cite{Klein2006} is often used to obtain starting values for the model parameters before applying iterative methods such as output-error \cite{Klein2006}, which is a maximum likelihood estimator for the problem where process noise is neglected. In other cases, measured states can be substituted in the first iteration of output-error so that initial parameter estimates are not needed. Using either the equation-error approach, or substituting the measured states in the first iteration, followed by application of output-error, can be also viewed as two-stage approaches.

\subsection{Contribution}

There are two main contributions of this paper. First, we show that there is a class of nonlinear estimation problems which arise in aerospace engineering applications that often have the mathematical structure of Eq. (\ref{eq:canonical_structure}). Second, we exploit this nonlinear structure to develop an estimator which naturally leads to a procedure for selecting good initial conditions for a given problem and have comparable (and in some instances better) accuracy and convergence characteristics relative to other nonlinear estimators currently used in aerospace applications.  We present two illustrative scalar examples to show how this estimator is implemented. Finally, we use this estimator to solve the problem of calibrating a 5-hole Pitot tube in flight. This problem is difficult to solve with a single stage estimator due to the nonlinearity and non-zero wind condition.

\subsection{Paper Organization}

The remainder of this paper is organized as follows. Section \ref{Estimator} describes the proposed estimator. The description includes a detailed derivation of the estimator equations and error bounds.  In Section \ref{ScalarEx}, the estimator is used solve two simple examples. These examples are tutorial in nature and show how the estimator is implemented in practice and how its performance compares to other nonlinear estimators. Then, in Section \ref{aeroEx}, we use the estimator to solve the 5-hole Pitot tube calibration problem using flight test data collected from a UAV. Section \ref{Conclusion} provides a summary and concluding remarks.

\section{Estimator Formulation}\label{Estimator}

In this section, we formulate the two-stage estimator, which is the subject of this paper.  We start by noting that the general nonlinear measurement model with additive noise from estimation theory \cite{Stengel} can be written as follows:
\begin{align}
{\bf z}_k &=  {\bf h}_k\left({\bf x}_k, {\bf u}_k,\boldsymbol{\xi}\right) + \bf{v}_k 
\end{align}
Without loss of generality, we are posing this as a parameter estimation problem.  As such, we have separated the parameters to be estimated, ${\bs \xi}$, from the states of the system ${\bf x}_k$.  We assume that this measurement model can be recast (as shown by the canonical form and examples in the appendices) into the form given by Eq. (\ref{eq:canonical_structure}) or:
\begin{align}
	{\bf z}_k &= {\bf A}\left({\bf x}_k, {\bf u}_k, {\bs \xi}_2\right){\bs \xi}_1 + {\bf b}\left({\bf x}_k, {\bf u}_k,{\bs \xi}_2\right) + {\bf v}_k, &E\left\{{\bf v}_k\right\} &= 0, 	&E\left\{{\bf v}_k{\bf v}_k^T\right\} &= {\bf R}
\label{eq:canonical_structure_2}	
\end{align}
where we assume ${\bf u}_k$ and ${\bf x}_k$ for $k = 1,...N$ are known. 

The measurement noise ${\bf v}_k$ is assumed to be independent, identically-distributed Gaussian white noise. Thus the covariance matrix ${\bf R}$ is set to be diagonal and its entries are unknown. As noted earlier, the algorithm proposed in this paper exploits the structure of Eq. (\ref{eq:canonical_structure_2}) as follows:  First, we solve a linear least squares problem for the parameter vector ${\bs \xi}_1$ where the remainder of the unknowns in the parameter vector ${\bs \xi}_2$ are held fixed at some appropriate and fixed values.  The algorithm includes a method for assessing the appropriateness of candidate ${\bs \xi}_2$ values. This is called the \emph{first stage}.  In the following \emph{second stage}, we solve a constrained nonlinear optimization problem for either ${\bs \xi}_2$ only (${\bs \xi}_1$ can be subsequently determined), or else for all of the unknowns (${\bs \xi}_1$ and ${\bs \xi}_2$) simultaneously, by using the estimates from the first-stage as the initial condition for the optimization. The choice of re-estimating either ${\bs \xi}_2$ only or else all the parameters in the second stage depends on the uniqueness of the local minimum. The determination is made empirically by a residual sampling procedure. This formulation leads to excellent convergence properties and, in many instances, guaranteed error bounds on the total cost function to be minimized.  It should be noted that this is different from the two-step estimator proposed by Haupt and Kasdin \cite{Haupt1996} in two fundamental ways.  First, a change of variables is not required.  Rather, the inherent structure of the problem is used in the two-stage process.  Second, the Haupt/Kasdin estimator uses estimates from their first-step process (a linear problem) as measurements in the second-step process (nonlinear optimization).  In the algorithm proposed here, the parameters are all estimated without having to formulate a pseudo-measurement by a change of variables.

To show why the proposed estimator works, we start by noting that the optimal estimate of the parameter vector ${\bs \xi}^{*}$ is the minimizer of the quadratic cost function $J({\bs \xi})$ with a penalized term on the covariance noise matrix ${\bf R}$, which is nonlinear in ${\bs \xi}$ and given below:
\begin{align}
{\bs \xi}^* &= \argmin_{\boldsymbol{\xi} \in \boldsymbol{\xi}_{limit}} J(\boldsymbol{\xi})\\
J(\boldsymbol{\xi}) & = J(\boldsymbol{\xi_1},\boldsymbol{\xi_2}) = \frac{1}{2}\sum_{k = 1}^{N} \w2norm({{\bf z}_k-{\bf A}\left({\bs \xi}_2\right){\bs \xi}_1 - {\bf b}\left({\bs \xi}_2\right)},{\bf R}) + \frac{N}{2} \ln{\vert{\bf R}\vert}
\label{costFunction2}
\end{align}
where we drop ${\bf x}_k$ and ${\bf u}_k$ from ${\bf A}\left({\bf x}_k, {\bf u}_k, {\bs \xi}_2\right)$ and ${\bf b}\left({\bf x}_k, {\bf u}_k, {\bs \xi}_2\right)$ to simplify the notation. $\boldsymbol{\xi}_{limit}$ is the constraint that is imposed on ${\bs \xi}$. This cost function is essentially the maximum likelihood estimation without the constant term \cite{Kay1993,Klein2006}. From Eq. (\ref{costFunction2}), it is clear that for a given, fixed value of $\boldsymbol{\xi_2}$ (which implies ${\bf A}\left({\bs \xi}_2\right)$ and ${\bf b}\left({\bs \xi}_2\right)$ are known), solving for $\boldsymbol{\xi_1}$ is nothing more than the traditional, linear least squares estimation problem if ${\bf R}$ is an identity matrix. Assuming  ${\bf R}$  is known for now (how the unknown ${\bf R}$ is handled is discussed in Sec. \ref{SelectionXi_2p}), the accuracy of the estimate for ${\bs \xi}_1$, denoted as $\hat{\boldsymbol{\xi}}_{1}$, will depend on how accurate ${\bf A}({\bs \xi}_2)$ is.  This, in turn, depends on how close a particular $\boldsymbol{\xi_2}$ used to form ${\bf A}({\bs \xi}_2)$, denoted as $\boldsymbol{\xi}_{2p}$, is to the optimal ${\bs \xi}^*_2$. If the initial guess ${\bs \xi}_{2p}$ is equal to $\boldsymbol{\xi^*_2}$, then the estimate of  $\boldsymbol{\xi_1}$ resulting from the linear least squares problem will be optimal.  However, since ${\bs \xi}^*_2$ is not known, how can we decide whether a given value of $\boldsymbol{\xi}_{2p}$ is close to ${\bs \xi}^*_2$?  We will answer this question by showing that the following are true:
\begin{enumerate}
\item The minimum of the cost function $J\left({\bs \xi}\right)$ is bounded from above and below by the error term $E$ ($E$ will be discussed in detail in the following Sec. \ref{bounding1} and \ref{bounding2})
	\begin{equation}
		J({\bs \xi}_1^*,{\bs \xi}_{2p})- E \leq J({\bs \xi}_1^*,{\bs \xi}_2^*) \leq J({\bs \xi}_1^*,{\bs \xi}_{2p})
		\label{eq:cost_function_bounds}
	\end{equation}
\item If ${\bf A}\left({\bs \xi}_2\right)$ and ${\bf b}\left({\bs \xi}_2\right)$ satisfy the Lipschitz condition and the domain of the state vector ${\bs \xi}$ is finite, then the cost function error $E$ is bounded.  Furthermore, the error term $E$ is a function of ${\bs \xi}_{2p}$.  
\end{enumerate} 

We will use these two points to develop a metric for assessing how close $J({\bs \xi}^{*}_1, {\bs \xi}_{2p})$ is to $J\left({\bs \xi}^{*}_1, {\bs \xi}^{*}_{2}\right)$.  This will be used to guide our selection of ${\bs \xi}_{2p}$ which will bring the cost function value in the first stage close to its optimal value.  Once we are close enough to the minimum value of $J\left({\bs \xi}_1, {\bs \xi}_2\right)$, we carry out the second stage optimization either on ${\bs \xi}_2$ only, or else on ${\bs \xi}_1$ and ${\bs \xi}_2$ simultaneously. The choice of determining whether to estimate one set or both sets can be empirically assessed by estimating trace of ${\bf R}$, denoted as $\mbox{Tr}[{\bf R}]$, in the first stage. If the estimated $\mbox{Tr}[{\bf R}]$ computed from a range of ${\bs \xi}_{2p}$ has a unique local minimum, then only ${\bs \xi}_2$ needs to be re-estimated. Otherwise, both ${\bs \xi}_1$ and ${\bs \xi}_2$ must be re-estimated simultaneously because the constraints for ${\bs \xi}_1$ and ${\bs \xi}_2$ in the sequential optimizing setting may not be valid. This is will be explained further in Sec. \ref{SelectionXi_2p}.  

It is observed that in some aerospace parameter estimation problems that ${\bs \xi}_2$ can be set to zero initially because it normally represents terms that are small biases or scale factor errors (c.f. Appendix A), and they are close to zero if the sensors are accurate. This information can also help determine ${\bs \xi}_{2p}$ qualitatively in addition to the quantitative procedure described in Sec. \ref{SelectionXi_2p}. In the next section, we show why the two points noted above are true.

\subsection{Bounding $J\left({\bs \xi}_1,{\bs \xi}_2\right)$} \label{bounding1}
To show that Eq. (\ref{eq:cost_function_bounds}) is true, we expand the cost function in Eq. (\ref{costFunction2}) as follows: 
\begin{equation}
\begin{aligned}
J({\bs \xi}_1,{\bs \xi}_2) &= \frac{1}{2}\sum_{k = 1}^{N} \w2norm({{\bf z}_k-{\bf A}\left({\bs \xi}_2\right){\bs \xi}_1 - {\bf b}\left({\bs \xi}_2\right)},{\bf R}) + \frac{N}{2} \ln{\vert{\bf R}\vert} \\
&= \frac{1}{2}\sum_{k = 1}^{N} \w2norm({{\bf z}_k-{\bf A}\left({\bs \xi}_{2p}\right){\bs \xi}_1 - {\bf b}\left({\bs \xi}_{2p}\right) - \left[{\bf A}\left({\bs \xi}_2\right) - {\bf A}\left({\bs \xi}_{2p}\right)\right]{\bs \xi}_1  - \left[{\bs b}\left({\bs\xi}_2\right) - {\bf b}\left({\bs \xi}_{2p}\right)\right]},{\bf R}) + \frac{N}{2} \ln{\vert{\bf R}\vert} \\
&\geq \underbrace{\frac{1}{2}\sum_{k = 1}^{N} \Bigg( \w2norm({{\bf z}_k-{\bf A}\left({\bs \xi}_{2p}\right){\bs \xi}_1 - {\bf b}\left({\bs \xi}_{2p}\right)},{\bf R})-\w2norm({\left[{\bf A}\left({\bs \xi}_2\right) - {\bf A}\left({\bs \xi}_{2p}\right)\right]{\bs \xi}_1},{\bf R}) -\w2norm({{\bf b}\left({\bs \xi}_2\right) - {\bs b}\left({\bs \xi}_{2p}\right)},{\bf R}) \Bigg) + \frac{N}{2} \ln{\vert{\bf R}\vert}}_{H({\bs \xi}_1,{\bs \xi}_2)}
\end{aligned}
\label{inequality1}
\end{equation}
The last inequality is obtained using the triangle inequality: $\left\Vert v + w\right\Vert \geq \left\Vert v \right\Vert - \left\Vert w \right\Vert $.

Thus, if we minimize both sides of Eq. (\ref{inequality1}), the following is obtained: 
\begin{equation}
\begin{aligned}
J^* &\geq \min_{\boldsymbol{\xi}_1,\boldsymbol{\xi}_2}
H(\boldsymbol{\xi}_1,\boldsymbol{\xi}_2) \\ 
&= \underbrace{\min_{\boldsymbol{\xi}_1,\boldsymbol{\xi}_2} \frac{1}{2}\sum_{k = 1}^{N} \w2norm({{\bf z}_k-{\bf A}\left({\bs \xi}_{2p}\right){\bs \xi}_1 - {\bf b}\left({\bs \xi}_{2p}\right)},{\bf R}) + \frac{N}{2} \ln{\vert{\bf R}\vert}}_{J\left(\boldsymbol{\xi}_1^*,\boldsymbol{\xi}_{2p}\right)} - \underbrace{\max_{\boldsymbol{\xi}_1,\boldsymbol{\xi}_2} \frac{1}{2}\sum_{k = 1}^{N} \left(\w2norm({\left[{\bf A}\left({\bs \xi}_2\right) - {\bf A}\left({\bs \xi}_{2p}\right)\right]{\bs \xi}_1},{\bf R}) +\w2norm({{\bf b}\left({\bs \xi}_2\right) - {\bf b}\left({\bs \xi}_{2p}\right)},{\bf R}) \right)}_{E} 
\end{aligned}
\label{eq:cost_function_lower_bound}
\end{equation} 
From the equation above, we see that $E$ is the error between the global optimal cost $J^*$ and the minimum of the first stage cost $J\left(\boldsymbol{\xi}_1^*,\boldsymbol{\xi}_{2p}\right)$ using a particular $\boldsymbol{\xi}_{2p}$.

By \emph{definition} of the optimum cost, the following inequality is true:

\begin{equation}
J^* \triangleq J\left(\boldsymbol{\xi}_1^*,\boldsymbol{\xi}_2^*\right) \leq J\left(\boldsymbol{\xi}_1^*,\boldsymbol{\xi}_{2p}\right)
\label{upperBound}
\end{equation}

Equation (\ref{eq:cost_function_bounds}) follows naturally from Eq. (\ref{eq:cost_function_lower_bound}) and (\ref{upperBound}).  It should be noted that Eq. (\ref{eq:cost_function_bounds}) \emph{does not} imply that there is a value of ${\bs \xi}_2 = {\bs \xi}_{2p}^{'}$ such that $J\left({\bs \xi}_1^{*},{\bs \xi}_{2p}^{'}\right) = J(\boldsymbol{\xi}_1^{*},\boldsymbol{\xi}_{2p})- E < J^*$.  Recall that in this first stage, we are selecting a value for ${\bs \xi}_2$ \emph{a priori} and the free variable is ${\bs \xi}_1$.  So for every value of ${\bs \xi}_2$ we select, the cost function for ${\bs \xi}_1$ changes.  Instead, the point articulated by Eq. (\ref{eq:cost_function_lower_bound}) is this: If the cost error $E$ is small, then $J(\boldsymbol{\xi}_1^*,\boldsymbol{\xi}_{2p}) \approx J^*$ and the result of the first stage cost is very close to the true optimal cost. In other words, the second step is now just a fine tuning of the first stage. In the next section, we derive bounds for the cost error $E$.

\subsection{Bounding $E = E\left({\bs \xi}_{2p}\right)$} \label{bounding2}
In general, it would be difficult to bound $E$ unless we place some restrictions on the nature of the functions ${\bf A}\left({\bs \xi}_2\right)$ and ${\bf b}\left({\bs \xi}_2\right)$ as well as the state vector ${\bs \xi} = \left[\begin{array}{cc} {\bs \xi}_1^T & {\bs \xi}_2^T \end{array}\right]^T$.  Thus, we will assume the following conditions hold true:
\begin{enumerate}
	\item The norm of the unknown parameter $\boldsymbol{\xi}_1$ is bounded by $\ell_1$: $\left\Vert\boldsymbol{\xi}_1 \right\Vert \leq \ell_1$
	\item The norm of the difference between  $\boldsymbol{\xi}_{2p}$ and $\boldsymbol{\xi}_{2}^{*}$ is bounded by $\ell_2$: $\left\Vert \boldsymbol{\xi}_{2}^* - \boldsymbol{\xi}_{2p} \right\Vert \leq \ell_2$
	\item The nonlinear function ${\bf A}\left({\bs \xi}_2\right)$ and ${\bf b}\left({\bs \xi}_2\right)$ are Lipschitz continuous functions and they satisfy the following: $\left\Vert
	{\bf A}\left({\bs \xi}^*_2\right) -{\bf A}\left({\bs \xi}_{2p}\right)\right\Vert \leq  L_{\boldsymbol{A}} \left\Vert {\bs \xi}^*_2 -{\bs \xi}_{2p} \right\Vert $ and  $ \left\Vert {\bf b}\left({\bs \xi}^*_2\right) -{\bf b}\left({\bs \xi}_{2p}\right)\right\Vert \leq L_{\boldsymbol{b}} \left\Vert {\bs \xi}^*_2 -{\bs \xi}_{2p}\right\Vert$ for ${\bs \xi}^*_2 < {\bs \xi}_2 < {\bs \xi}_{2p}$
\end{enumerate}
where $\ell_1$ and $\ell_2$ are scalars, and $L_{\boldsymbol{A}}$ and $L_{\boldsymbol{b}}$ are called Lipschitz constants (also scalars).  The first two conditions are satisfied if the state vector ${\bs \xi}$ has a finite domain. This is a reasonable assumption in many engineering problems where the state vector represents some physical and measurable quantity.  The upper bound $\ell_1$ in the first assumption represents the maximum value that $\boldsymbol{\xi}_1$ can achieve. The upper bound  $\ell_2$ in the second assumption represents the error between the initial guess and optimal value of $\boldsymbol{\xi}_2$.  Thus, these two conditions are not very restrictive. The values of $\ell_1$ and $\ell_2$ can be usually estimated based on the prior knowledge. For example, the absolute value of a reasonable scale factor ${\bs \xi}_{2}$ should not be bigger than 0.5 (i.e. $-0.5 \leq {\bs \xi}^*_{2} \leq 0.5$ and this bound is very conservative). Then we can pick ${\bs \xi}_{2p}$ such that  $|{\bs \xi}^*_{2} - {\bs \xi}_{2p}| \leq 0.5 + |{\bs \xi}_{2p}| \leq  \ell_2 $. ${\bs \xi}_{2p}$ should be chosen such that it is close to ${\bs \xi}^*_{2}$. If ${\bs \xi}_{2p}$ is set to be 1, then  $\ell_2 $ can be set to 1.5 to upper bound $|{\bs \xi}^*_{2} - {\bs \xi}_{2p}|$. The third condition requiring the functions $\mathbf{A}\left({\bs \xi}_2\right)$ and $\mathbf{b}\left({\bs \xi}_2\right)$ to be Lipschitz continuous is not very restrictive either.  Many mathematical functions used to model physical systems, such as the square root (real positive numbers under the square root), as well as sine and cosine functions, are Lipschitz continuous. Furthermore, $L_{\boldsymbol{A}}$ and $L_{\boldsymbol{b}}$ can also be viewed as the derivative information of  ${\bs \xi}_{2}$. If the selected ${\bs \xi}_{2p}$ approaches ${\bs \xi}^*_{2}$, then $L_{\boldsymbol{A}}$ and $L_{\boldsymbol{b}}$ approach zero. With these three assumptions, we can upper bound the following two error terms:

\begin{equation}
\begin{aligned}
E_1 &\triangleq \max_{{\bs \xi}_1,{\bs \xi}_2} \left\Vert \left[{\bf A}\left({\bs \xi}_2\right) -{\bf A}\left({\bs \xi}_{2p}\right) \right] {\bs \xi}_1 \right\Vert  \leq  \max_{{\bs \xi}_1,{\bs \xi}_2} \left\Vert {\bf A}\left({\bs \xi}_2\right) -{\bf A}\left({\bs \xi}_{2p}\right)  \right\Vert  \left\Vert  {\bs \xi}_1 \right\Vert  \\
& \leq  \max_{{\bs \xi}_2}  \left\Vert  {\bf A}\left({\bs \xi}_2\right) -{\bf A}\left({\bs \xi}_{2p}\right)  \right\Vert  \ell_1 \leq  L_{\boldsymbol{A}} \left\Vert {\bs \xi}^*_2 -{\bs \xi}_{2p}   \right\Vert   \ell_1  \\
&\leq  L_{\boldsymbol{A}} \ell_2   \ell_1  \\
\end{aligned} 
\label{errorBound2}
\end{equation}

\begin{equation}
\begin{aligned}
E_2 &\triangleq  \max_{{\bs \xi}_1,{\bs \xi}_2} \left\Vert {\bf b}\left({\bs \xi}_2\right) -{\bf b}\left({\bs \xi}_{2p}\right)  \right\Vert \leq  L_{\boldsymbol{b}} \left\Vert  {\bs \xi}^*_2 -{\bs \xi}_{2p}   \right\Vert \leq   L_{\boldsymbol{b}} \ell_2  \\
\end{aligned} 
\label{errorBound3}
\end{equation}
where the first inequality in Eq. (\ref{errorBound2}) comes from Cauchy-Schwarz inequality. Using Eq. (\ref{errorBound2}) and (\ref{errorBound3}) we can derive an upper bound on the error $E$ as follows:

\begin{equation}
\begin{aligned}
E &= \max_{\boldsymbol{\xi}_1,\boldsymbol{\xi}_2} \frac{1}{2}\sum_{k = 1}^{N} \left(\w2norm({\left[{\bf A}\left({\bs \xi}_2\right) - {\bf A}\left({\bs \xi}_{2p}\right)\right]{\bs \xi}_1},{}) +\w2norm({{\bf b}\left({\bs \xi}_2\right) - {\bf b}\left({\bs \xi}_{2p}\right)},{}) \right) \\
  & =\frac{1}{2}\sum_{k = 1}^{N} \left(\max_{\boldsymbol{\xi}_1,\boldsymbol{\xi}_2} \left\Vert{\left[{\bf A}\left({\bs \xi}_2\right) - {\bf A}\left({\bs \xi}_{2p}\right)\right]{\bs \xi}_1}\right\Vert \right)^2 + \frac{1}{2}\sum_{k = 1}^{N} \left(\max_{\boldsymbol{\xi}_1,\boldsymbol{\xi}_2} \left\Vert{{\bf b}\left({\bs \xi}_2\right) - {\bf b}\left({\bs \xi}_{2p}\right)}\right\Vert \right)^2 \\
&=\frac{1}{2}\sum_{k = 1}^{N} (E_1^2  + E_2^2) \leq \frac{N}{2} \big(  L_{\boldsymbol{A}}^2 \ell_2^2   \ell_1^2  +   L_{\boldsymbol{b}}^2 \ell_2^2  \big) = \frac{N}{2} \ell_2^2 \big(  L_{\boldsymbol{A}}^2  \ell_1^2  +   L_{\boldsymbol{b}}^2   \big)  \\
\end{aligned} 
\label{errorBoundtot}
\end{equation}
where we dropped the subscript ${\bf R}$ without loss of generality. Equation (\ref{errorBoundtot}) implies that for a fixed length of data set $N$, if the initial guess ${\bs \xi}_{2p}$ is close to the optimal ${\bs \xi}^*_2$ (i.e., $L_{\boldsymbol{A}}$ and $L_{\boldsymbol{b}}$ approach zero) and the bounds are ${\bs \xi}_{1}$  and ${\bs \xi}_{2}$ are small (i.e., $\ell_1$ and $\ell_2$ approach zero),  then the first stage optimization cost function is close to the original cost function (i.e., $E$ consequently approaches zero). This means that the result of the first stage can bring the cost very close to the minimum global cost, which makes the second stage more likely to converge. 

\subsection{Selection of ${\bs \xi}_{2p}$}\label{SelectionXi_2p}
So how do we select ${\bf \xi}_{2p}$ so that $E$ is small, thereby assuring that the second stage optimization will lead to the correct solution?  While it is difficult to develop a prescriptive solution for selecting ${\bf \xi}_{2p}$, we can answer the following related question:  How do we know if a given choice of ${\bs \xi}_{2p}$ is one that will increase the chances of convergence to the correct solution?  To answer this question, we start by linearizing ${\bf A}\left({\bs \xi}_2\right)$ and ${\bf b}\left({\bs \xi}_2\right)$ with respect to ${\bs \xi}_2$ at ${\bs \xi}_{2p}$ as follows:
\begin{equation}
\begin{aligned}
{\bf A}\left({\bs \xi}_2\right) &\approx {\bf A}\left({\bs \xi}_{2p}\right) + \frac{\partial {\bf A}\left({\bs \xi}_{2p}\right)}{\partial {\bs \xi}_2}  \Delta {\bs \xi}_2 \\
{\bf b}\left({\bs \xi}_2\right) &\approx {\bf b}\left({\bs \xi}_{2p}\right) + \frac{\partial {\bf b}\left({\bs \xi}_{2p}\right)}{\partial {\bs \xi}_2} \Delta {\bs \xi}_2\\
\end{aligned}
\label{linearization}
\end{equation}

If ${\bs \xi}_{2p}$ is chosen such that the first order terms are sufficient small and satisfy Eq. (\ref{taylorExpansion}),

\begin{equation}
\begin{aligned}
&\left\Vert \frac{\partial {\bf A}\left({\bs \xi}_{2p}\right)}{\partial {\bs \xi}_2}  \Delta {\bs \xi}_2  \right\Vert \le \left\Vert \frac{\partial {\bf A}\left({\bs \xi}_{2p}\right)}{\partial {\bs \xi}_2}\right\Vert \ell_2 \ll 	 \left\Vert {\bf A}\left({\bs \xi}_{2p}\right) \right\Vert \\
&\left\Vert \frac{\partial {\bf b}\left({\bs \xi}_{2p}\right)}{\partial {\bs \xi}_2} \Delta {\bs \xi}_2 \right\Vert \le \left\Vert \frac{\partial {\bf b}\left({\bs \xi}_{2p}\right)}{\partial {\bs \xi}_2} \right\Vert \ell_2 \ll \left\Vert{\bf b}\left({\bs \xi}_{2p}\right) \right\Vert
\end{aligned}
\label{taylorExpansion}
\end{equation}
then the nonlinear parameter cost function in Eq. (\ref{costFunction2}) can be approximated by:

\begin{equation}
\begin{aligned}
&\argmin_{{\bs \xi}_1,{\bs \xi}_2}  \frac{1}{2}\sum_{k = 1}^{N} \w2norm({{\bf z}_k-{\bf A}\left({\bs \xi}_2\right){\bs \xi}_1 - {\bf b}\left({\bs \xi}_2\right)},{\bf R}) + \frac{N}{2} \ln{\vert{\bf R}\vert}  \\
\approx &\argmin_{{\bs \xi}_1} \frac{1}{2} \sum_{k = 1}^{N} \left\Vert{\bf z}_k-{\bf A}\left({\bs \xi}_{2p}\right){\bs \xi}_1 - {\bf b}\left({\bs \xi}_{2p}\right) \right\Vert^2_{{\bf R}({\bs \xi}_{2p})} + \frac{N}{2} \ln{\vert{\bf R}({\bs \xi}_{2p})\vert} \\
\end{aligned} 
\label{costNew}
\end{equation}
where ${{\bf R}({\bs \xi}_{2p})}$ is still unknown but it is a matrix that depends on  ${\bs \xi}_{2p}$. Equation (\ref{costNew}) implies that the linearized system cost is close to the original nonlinear system cost. The right-hand side of Eq. (\ref{costNew}) can be solved using linear least squares by setting the unknown ${{\bf R}({\bs \xi}_{2p})}$ equal to the identity matrix.  Note that ${\bs \xi}_{1p}$ is suboptimal (biased) in the first stage due to the unknown ${\bf R}$ and  the bounding properties shown in Eq. (\ref{eq:cost_function_bounds}) and (\ref{errorBoundtot})  do not change except $J\left(\boldsymbol{\xi}_1^*,\boldsymbol{\xi}_{2p}\right) = J\left(\boldsymbol{\xi}_{1p},\boldsymbol{\xi}_{2p}\right)$. The unknown ${\bf R}$ and ${\bs \xi}_{2}$ are solved optimally via the second stage nonlinear optimization. 

There may be one or more suboptimal pairs (${\bs \xi}_{1p}$,${\bs \xi}_{2p}$) obtained from solving the linearized system, that has a cost value approximately equal to the optimal cost. Since estimating parameters using linear least square is not computationally expensive, we can sample a large pool of ${\bs \xi}_{2p}$ from the feasible set (constrained by $\left\Vert \boldsymbol{\xi}_{2}^* - \boldsymbol{\xi}_{2p} \right\Vert \leq \ell_2$ ) to estimate the suboptimal ${\bs \xi}_{1p}$. Also, the parameter ${\bs \xi}_{2p}$ should satisfy Eq. (\ref{theta2p}) and (\ref{secondOrder}):

\begin{equation}
\begin{aligned}
\frac{\left\Vert \dfrac{\partial {\bf A}\left({\bs \xi}_{2p}\right)}{\partial {\bs \xi}_2}  \right\Vert \ell_2}{ \left\Vert {\bf A}\left({\bs \xi}_{2p}\right) \right\Vert}  \leq T_1 \quad \text{and} \quad
\frac{\left\Vert \dfrac{\partial {\bf b}\left({\bs \xi}_{2p}\right)}{\partial {\bs \xi}_2}  \right\Vert \ell_2}{ \left\Vert{\bf b}\left({\bs \xi}_{2p}\right) \right\Vert} \leq T_2 
\end{aligned}
\label{theta2p}
\end{equation}

\begin{equation}
\sum_{i = 1}^{m_{\bf A}}\sum_{j = 1}^{n_{\bf A}}\frac{\partial {\bf A}(i,j)^2\left({\bs \xi}_{2p}\right)}{\partial {\bs \xi}_2\partial {\bs \xi}^T_2}  > {\bf 0}_{n_{{\bs \xi}_2} \times n_{{\bs \xi}_2}} \quad \mbox{and} \quad \sum_{i = 1}^{m_{\bf b}} \frac{\partial {\bf b}(i)^2\left({\bs \xi}_{2p}\right)}{\partial {\bs \xi}_2\partial {\bs \xi}^T_2}  > {\bf 0}_{n_{{\bs \xi}_2} \times n_{{\bs \xi}_2}}
\label{secondOrder}
\end{equation}
where $T_1$ and $T_2$ are user-defined and can be interpreted as percentage requirements, and $n_{{\bs \xi}_2}$ is the number of parameters in ${\bs \xi}_2$. The smaller the values of $T_1$ and $T_2$ (obtained through varying ${\bs \xi}_{2p}$), the tighter the error bound on $E$. Equation (\ref{theta2p}) ensures validity of linearization in the first stage and Eq. (\ref{secondOrder}) enforces local convergence for iterative methods in the second stage.   

Once we have chosen a set of ${\bs \xi}_{2p}$, we can estimate the residual vector history ${\bs v}_k$ for $k =1...N$ and use it to build a metric to find a suitable pair (${\bs \xi}_{1p}$,${\bs \xi}_{2p}$) for the second stage nonlinear estimation. Namely, we find the suboptimal pair ($\hat{{\bs \xi}}_{1p}$,$\hat{{\bs \xi}}_{2p}$) by solving  Eq. (\ref{costNew2}): 

\begin{equation}
\begingroup 
\setlength\arraycolsep{1.5pt}
\renewcommand{\arraystretch}{0.7}
\begin{aligned}
\argmin_{{\bs \xi}_{2p}} \mbox{Tr} \big[{\bf R}({\bs \xi}_{2p})\big] = &\argmin_{{\bs \xi}_{2p}}  \sum_{k = 1}^{N} \mbox{Tr} \big({\bs v}_k{\bs v}^T_k \big)\\
 &\mbox{where} \: {\bs v}_k =  {\bf z}_k-{\bf A}({\bs \xi}_{2p})\hat{{\bs \xi}}_{1p} - {\bf b}({\bs \xi}_{2p}) \\
 &\mbox{and} \: \hat{{\bs \xi}}_{1p} = (\mathcal{A}^T\mathcal{A})^{-1}\mathcal{A}^T({\bf Z} - \mathcal{B}) \\
 &\mbox{for} \: {\bs \xi}_{2p} \in S_{{\bs \xi}_{2}}
\end{aligned}
\endgroup
\label{costNew2}
\end{equation}
where $\mathcal{A}$, $\mathcal{B}$ and ${\bf Z}$ are concatenations of ${\bf A}_k({\bs \xi}_{2p})$, ${\bf b}_k({\bs \xi}_{2p})$ and ${\bf z}_k$ respectively for $k = 1,...,N$. $S_{{\bs \xi}_{2}}$ is a chosen set that satisfies the constraint $\left\Vert \boldsymbol{\xi}_{2}^* - \boldsymbol{\xi}_{2p} \right\Vert \leq \ell_2$ and Eq. (\ref{theta2p}) and (\ref{secondOrder}).  
By minimizing the trace of ${\bf R}({\bs \xi}_{2p})$, we are essentially finding the suboptimal pair that gives the smallest residual vector. We denote this method as the residual sampling procedure. Note that $\mbox{Tr} \big[{\bf R}({\bs \xi}_{2p})\big]$ is a similar measure of the error term $E$ shown in Eq. (\ref{eq:cost_function_bounds}), where $E$ can be interpreted as a weighted residual least squares error.
    
If the estimated $\mbox{Tr} \big[{\bf R}({\bs \xi}_{2p})\big]$ has a local minimum, then only ${\bs \xi}_2$ needs to be re-estimated in the second stage. Estimating only ${\bs \xi}_2$ also means the search space in the nonlinear programming is significantly reduced. Once ${\bs \xi}^*_2$ and ${\bf R}$ are estimated alternately in the second stage, ${\bs \xi}^*_1$ is immediately calculated using weighted linear least squares. We also use ${\bf R}({\bs \xi}_{2p})$ to initialize ${\bf R}$ in the second stage, as shown in Eq. (\ref{costNew}). If the estimated  $\mbox{Tr} \big[{\bf R}({\bs \xi}_{2p})\big]$  does not have a unique local minimum (as shown in Sec. \ref{aeroEx}), both ${\bs \xi}_1$ and ${\bs \xi}_2$ should be re-estimated simultaneously in the second stage. This is because the sequential order of constraints may not be valid. Namely, 

\begin{equation}
\min_{{\bs \xi}_1 \in S_{{\bs \xi}_{1}},{\bs \xi}_2 \in S_{{\bs \xi}_{2}}} J({\bs \xi}_1,{\bs \xi}_2) \neq \min_{{\bs \xi}_2 \in S_{{\bs \xi}_{2}}}\bigg[\min_{{\bs \xi}_1 \in S_{{\bs \xi}_{1}}} J({\bs \xi}_1,{\bs \xi}_2)\bigg]
\label{equalivant}
\end{equation} 

When $\hat{{\bs \xi}}_1$ from the inner minimization on the right-hand side of Eq. (\ref{equalivant}) cannot be uniquely determined, the outer minimization may not be able to arrest $\hat{{\bs \xi}}_1$ escaping from its own constraint. Though this inequality holds true in general, we observe that if the inner minimization has a unique solution (i.e., the error $E$ is small) using a large sample of ${\bs \xi}_{2p}$ from the feasible set $S_{{\bs \xi}_{2}}$, then both sides of Eq. (\ref{equalivant}) can be equal. In other words, since the search space of ${\bs \xi}_{2}$ in the inner minimization has been searched exhaustively via sampling, the chance of $\hat{\bs \xi}_{1}$ escaping from the outer minimization is small. Therefore, if we cannot clearly find a unique local minimum in the first stage represented by the inner minimization, we need to re-estimate  ${\bs \xi}_1$ and ${\bs \xi}_2$ simultaneously by solving the left-hand side of Eq. (\ref{equalivant}).  The estimates $\hat{{\bs \xi}}_{1p}$ and  $\hat{{\bs \xi}}_{2p}$ from the first stage are still used as the initial condition, where $\hat{{\bs \xi}}_{2p}$ is any vector of the set that results in multiple local minima.   

Though this residual sampling method is very crude, it does provide an excellent initial condition for the second stage, as will be demonstrated by examples in Sections \ref{ScalarEx} and \ref{aeroEx}. One possible alternative of selecting ${\bs \xi}_{2p}$ would be to evaluate the Jacobian $\nabla J({\bs \xi}_{2p})$ and iteratively update ${\bs \xi}_{2p}$ until $\nabla J({\bs \xi}_{2p}) = 0$. However, this method is computationally expensive and prone to error when the nonlinear functions ${\bf A}({\bs \xi}_{2p})$ and ${\bf b}({\bs \xi}_{2p})$ are multi-dimensional and highly nonlinear. The effect of this selection of ${\bs \xi}_{2p}$ is depicted graphically in Fig. \ref{errorIteration}, where ${\bs \xi}^{(j)}_{2p}$ is a not good choice; it does not give the smallest  $\mbox{Tr} \big[{\bf R}({\bs \xi}_{2p})\big]$ and it may cause the second stage to arrive the wrong minimum even though it is within the bound of $\ell_2$.  On the other hand, ${\bs \xi}^{(i)}_{2p}$ is a good choice because 1) it is the local minimum in the constraint set $\ell_2$ and 2) the positive concavity (concave up) ensures local convergence. 

Putting all of this together results in the following procedure for implementation of the proposed algorithm:
\begin{figure}[!ht]
	\centering
	\includegraphics[width=.5\textwidth]{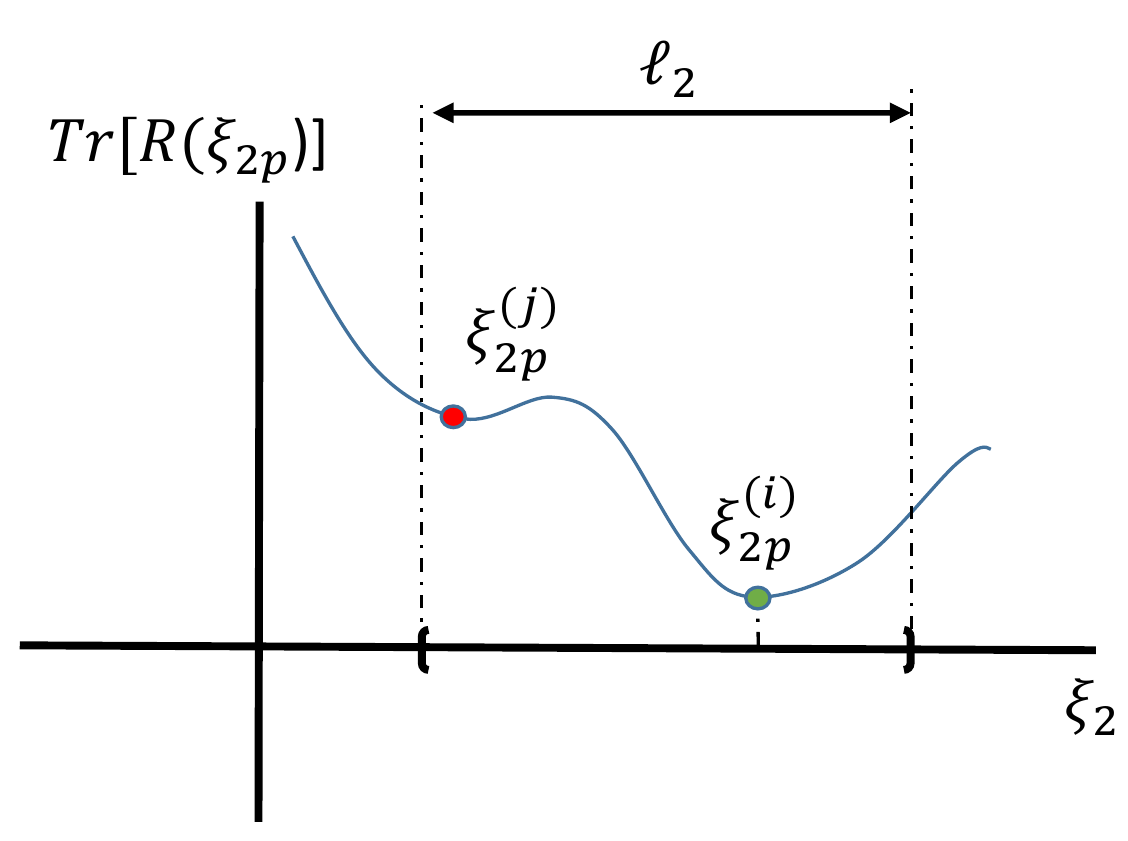}
	\caption{Pictorial depiction of the effect of choices of ${\bs \xi}_{2p}$ on  $\mbox{Tr} \big[{\bf R}({\bs \xi}_{2p})\big]$ value.}
	\label{errorIteration}
\end{figure}

\begin{enumerate}[label=Step \arabic*:]
	\item Formulate the measurement equation to have the form given by Eq. (\ref{eq:canonical_structure_2}).
	\item Sample a large pool of ${\bs \xi}_{2p}$ from the feasible constraint set $\left\Vert \boldsymbol{\xi}_{2}^* - \boldsymbol{\xi}_{2p} \right\Vert \leq \ell_2$; those ${\bs \xi}_{2p}$ should also satisfy Eq. (\ref{theta2p}) and (\ref{secondOrder}).
	\item Estimate ${\bs \xi}_{1p}$ by minimizing the cost function (right-hand side of Eq. \ref{costNew}) using linear least squares with the unknown ${\bf R} = {\bf I}$. Then calculate the corresponding trace $\mbox{Tr} \big[{\bf R}({\bs \xi}_{2p})\big]$.  
	\item Find a suboptimal pair ($\hat{{\bs \xi}}_{1p}$,${\bs \xi}_{2p}$) such that the corresponding $\mbox{Tr} \big[{\bf R}({\bs \xi}_{2p})\big]$ has a unique local minimum. If there exist multiple suboptimal pairs (similar numerical values), choose an arbitrary one from these suboptimal pairs. 
	This completes the first stage.  
	\item If there exist a unique local minimum from $\mbox{Tr} \big[{\bf R}({\bs \xi}_{2p})\big]$, then solve for ${\bs \xi}_2$ only in the second stage with ${\bs \xi}_{2p}$ as the initial condition. Use ${\bf R}({\bs \xi}_{2p})$ to initialize ${\bf R}$ in the second stage. Once ${\bs \xi}^*_2$ and ${\bf R}$ are obtained,  ${\bs \xi}^*_1$ immediately follows using weighted linear least squares. Otherwise, solve for both  ${\bs \xi}_{1}$ and ${\bs \xi}_2$ simultaneously with the suboptimal estimate ($\hat{{\bs \xi}}_{1p}$,${\bs \xi}_{2p}$) as the initial condition in the second stage. The nonlinear function can be minimized by any standard iterative method such as modified Newton-Raphson, Gauss-Newton or Levenberg-Marquardt \cite{Marquardt1963} method. We estimate ${\bs \xi}$ and  ${\bf R}$  alternately until both ${\bs \xi}$ and the diagonal elements of ${\bf R}$ converge. This completes the second stage. 	
\end{enumerate}

Though the measurement model in Eq. (\ref{eq:canonical_structure_2}) resembles a Kalman Filter (KF) or Extended Kalman filter (EKF) measurement model equation, we find that it is not straightforward to make the proposed algorithm a stand-alone measurement equation in a recursive estimation. This is because of the nature of the first stage, where the optimality and separability of $\hat{{\bs \xi}}_{1}$ depends on $\hat{{\bs \xi}}_{2}$ generally in a non-linear fashion. However, one can use the first stage of the proposed algorithm to estimate an initial condition with a small batch of data for an EKF or Iterated-EKF (IEKF) filter. Then we can use $\dot{\hat{{\bs \xi}}} = 0$ as the parameter time update equation and linearize the measurement in Eq. (\ref{eq:canonical_structure_2}) with respect to ${\bs \xi}$ to formulate the linearized measurement matrix needed for EKF or IEKF. This is demonstrated in Sec. IV of Ref. \cite{Crassidis2005}.  In the following section, we provide a demonstration on how to implement this estimator.

\section{Two Tutorial Examples: Scalar Measurement Equations}\label{ScalarEx}
To demonstrate the mechanics of using this estimator, gain some intuition into its operation and compare its performance to other estimators, we solve the following static parameter estimation problem which is a simplified version of the problem presented in Ref. \cite[Eq.~(36)]{Haupt1996}:
\begin{equation}
z_k = f_k(\eta_k) + v_k= \underbrace{(1+a)\cos(\eta_k + b) + c }_{f_k\left(\eta_k\right)} + v_k
\label{model1}
\end{equation}
The variables $a$, $b$ and $c$ (the coefficients of the nonlinear function $f_k)$ are the parameters we want to estimate. In this particular case,  we set the values of the parameters as follows: $a = 1$ , $b = 0.1$ and $c = 1$. There are 100 scalar measurements $z_k$ generated by varying $\eta$ from $1$ to $10$ radians, incrementing by the same interval. The 100 measurement noise $v_k$ is drawn from a normal distribution with mean of zero and a standard deviation of 0.3.

In order to use the estimator developed in this paper on Eq. (\ref{model1}), the scalar measurement model is recast into an affine problem by exploiting the structure of the nonlinear function $f_k\left(\eta_k\right)$ as shown below:
\begin{equation}
\begingroup 
\setlength\arraycolsep{1.5pt}
\renewcommand{\arraystretch}{0.7}
z_k =\underbrace{\left[\begin{array}{cc}
	\cos(\eta_k + b) & 1
	\end{array}\right]}_{{\bf A}\left({\bs \xi}_2\right)} \underbrace{\begin{bmatrix}
	a \\
	c \\
	\end{bmatrix}}_{{\bs \xi}_{1}}+ \underbrace{\cos(\eta_k + b)}_{{\bf b}\left({\bs \xi}_2\right)} + v_k
\endgroup
\label{method2}
\end{equation}
where ${\bs \xi}_{1} = [a,c]^T$ and ${\bs \xi}_2 = b$.

\begin{figure}[!ht]
	\centering
	\includegraphics[width=.5\textwidth]{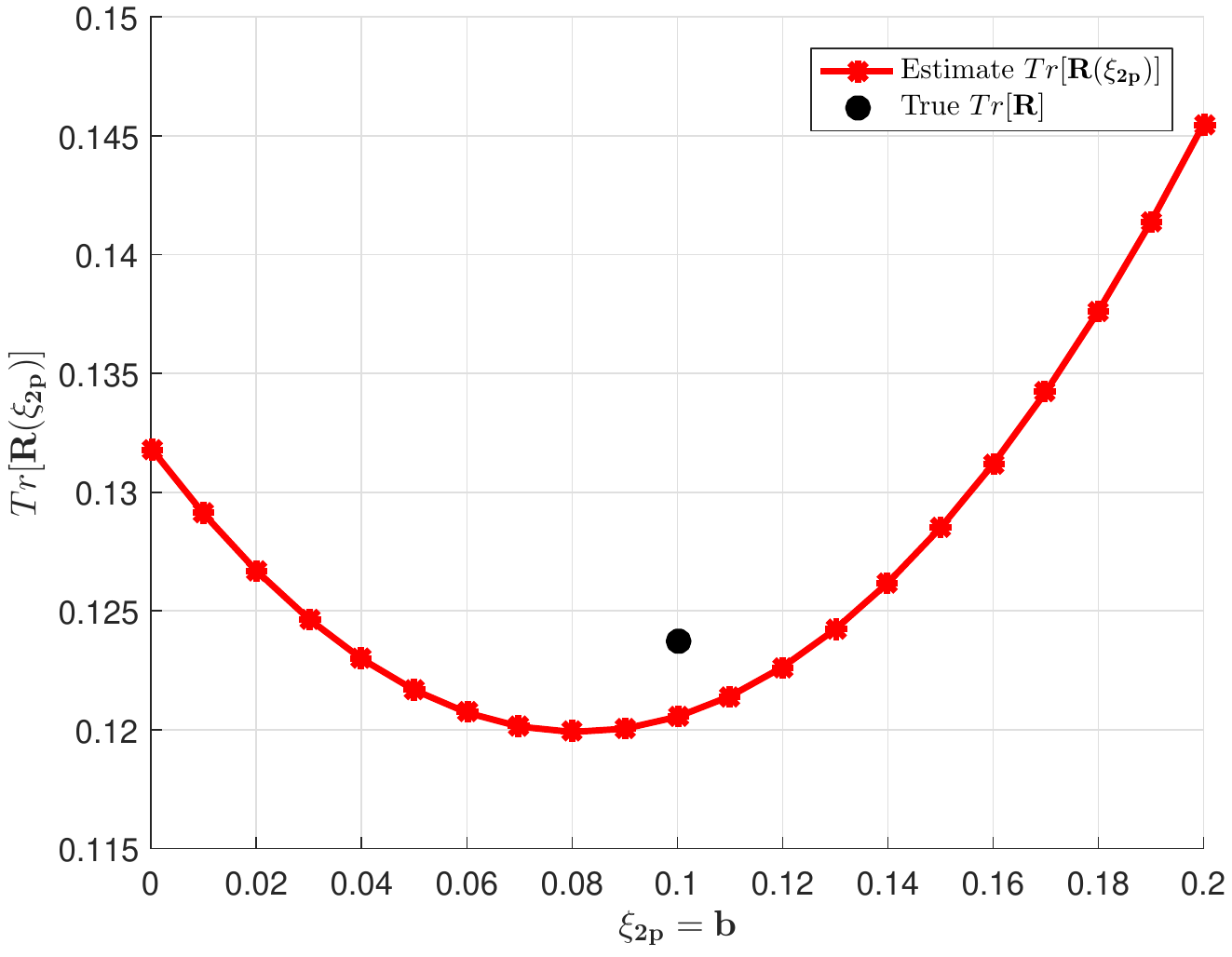}
	\caption{Estimated $\mbox{Tr} \big[{\bf R}({\bs \xi}_{2p})\big]$ by sampling random of $\xi_{2p}$ for scalar example 1.}
	\label{xi_2p_1_Ex}
\end{figure}

Since there is only one parameter in ${\bs \xi}_2$, we can simply sweep a range of $b$ to estimate $\mbox{Tr} \big[{\bf R}({\bs \xi}_{2p})\big]$. Also, $\mbox{Tr} \big[{\bf R}({\bs \xi}_{2p})\big] = {\bf R}({\bs \xi}_{2p}) $ for this problem since the measurement at each time step is a scalar. Figure \ref{xi_2p_1_Ex} shows the estimated scalar value of $\mbox{Tr} \big[{\bf R}({\bs \xi}_{2p})\big]$. It can be seen ${\bs \xi}_{2p} = b = 0.08$ corresponds the minimum value of ${\bf R}({\bs \xi}_{2p})$. We also observe that both $b = 0.08$ and its corresponding ${\bf R}(0.08)$ are not same as the true values due to the measurement noise. Nonetheless, there exists a unique minimum so we will use ${\bs \xi}_{2p} = 0.08$ to estimate ${\bs \xi}_{1p}$.

With the unique measurement structure, valid linear approximation (Eq. \ref{costNew})) and a unique local minimum (indicated by the positive definiteness of ${\bf A}({\bs \xi}_{2p})$ and ${\bf b}({\bs \xi}_{2p})$ (Eq. (\ref{secondOrder})), the parameters $a$ and $c$ are estimated in first stage by solving a linear least squares problem which minimizes the following cost function:
\begin{align}
	\hat{{\bs \xi}}_{1p} = \argmin_{\boldsymbol{\xi}_{1}}\bigg( \frac{1}{2}\sum_{k = 1}^{N} \w2norm({z_k-{\bf A}\left({0.08}\right){\bs \xi}_1 - {\bf b}\left({0.08}\right)},{\bf I})\bigg)
	\label{xi_1p}
\end{align}
Since there exists a unique minimum as shown in Fig. \ref{xi_2p_1_Ex},  we use  ${\bs \xi}_{2p} = 0.08$ as the initial condition to estimate ${\bs \xi}_{2}$ in the second stage:

\begin{equation}
\hat{{\bs \xi}}_{2}, \hat{\bf R} = \argmin_{\boldsymbol{\xi_{2}},{\bf R} } \frac{1}{2} \bigg([\bf{Z} - \mathcal{A}{\bs \xi}_1 - \mathcal{B} ]^T{\bf W}[\bf{Z} -\mathcal{A}{\bs \xi}_1 - \mathcal{B}] \bigg)
\label{method1}
\end{equation}
where 
\begin{equation}
\begingroup 
\setlength\arraycolsep{1.5pt}
\renewcommand{\arraystretch}{0.7}
{\bs \xi}_2 = 
b_p = 0.08 \hspace{0.2in}
\bf{Z} = \begin{bmatrix}
z_1 \\
\vdots \\
z_N
\end{bmatrix} \hspace{0.2in}
\mathcal{A} = \begin{bmatrix}
{\bf A}_1\left({\bs \xi}_2\right) \\
\vdots \\
{\bf A}_N\left({\bs \xi}_2\right)
\end{bmatrix}
\hspace{0.2in}
\mathcal{B} = \begin{bmatrix}
{\bf b}_1\left({\bs \xi}_2\right) \\
\vdots \\
{\bf b}_N\left({\bs \xi}_2\right)
\end{bmatrix}
\hspace{0.2in}
{\bf W} = \left[\begin{array}{ccc} {\bf R}^{-1} & & \\
& \ddots & \\ 
&   &  {\bf R}^{-1} \end{array}\right]
\endgroup
\label{method1_1}
\end{equation}
The sequential quadratic programming (SQP) algorithm is used to solve the optimization problem given by Eq. (\ref{method1}). Note that ${\bs \xi}_1$ is calculated iteratively using weighted linear least squares inside the nonlinear cost solver, so there is no need to initialize  ${\bs \xi}_1$ in the beginning of the second stage. The term ${\bs \xi}_{1p}$ in Eq. (\ref{xi_1p}) is used to initialize the second stage if the local minimum is not unique (demonstrated later in Sec. \ref{aeroEx}). For the work reported in this paper, the SQP is implemented using the built-in MATLAB function \texttt{fmincon} \cite{fmincon}. An outer while-loop outside of \texttt{fmincon} is written to estimate ${\bf R}$ alternately with ${\bs \xi}_{2}$ until the following is satisfied (Eq.(6.41e) in Ref. \cite{Klein2006}):  

\begin{equation}
\Bigg\vert \frac{(\hat{r}_{jj})_k - (\hat{r}_{jj})_{k-1} }{(\hat{r}_{jj})_{k-1}}\Bigg\vert < 0.05 \quad \forall j, \: j =1,2,...,n_o
\label{rrCon}
\end{equation}
where $(\hat{r}_{jj})$ is the estimate of the $j$th diagonal element of the estimate $\hat{\bf R}$ and $n_o$ is the number of the total diagonal terms. In this scalar example, $j=1$ since  ${\bf R}$ is a scalar.  Once the optimal $\hat{{\bs \xi}}^*_{2}$ and $\hat{\bf R}$ are obtained, $\hat{{\bs \xi}}^*_{1}$ can be immediately solved using weighted linear least squares:

\begin{equation}
\hat{{\bs \xi}}^*_{1} = (\mathcal{A}^T  {\bf W}\mathcal{A})^{-1}\mathcal{A}^T{\bf W}({\bf Z} - \mathcal{B}) 
\end{equation}
We will benchmark the performance of this estimator against the following pair of nonlinear estimators: (1) a classic, nonlinear program which solves for ${\bs \xi}_1$ and ${\bs \xi}_2$ simultaneously and (2) the Haupt/Kasdin two-step estimator described in Ref. \cite{Haupt1996}. 

\subsection{Benchmark 1: Classic Non-Linear Programming}\label{M1}
The first benchmark is nothing more than a solution to the optimization problem posed by the left-hand side of  Eq. (\ref{costNew}). The implementation of this benchmark differs from the algorithm proposed in this paper, since the initial conditions are selected randomly. 

\subsection{Benchmark 2: Haupt/Kasdin Two-Step Estimator}\label{M3}	
To implement the Haupt/Kasdin two-step estimator, we choose a new set of states by a change of variables such that Eq. (\ref{model1}) can be written as a linear measurement model shown below:
\begin{equation}
\begingroup 
\setlength\arraycolsep{1.5pt}
\renewcommand{\arraystretch}{0.7}
\begin{aligned}
z_k &=  {\bf H}_k {\bf f}({\bs \xi}) = {\bf H}_k{\bf y} + v_k \\
& = \underbrace{\begin{bmatrix}
\cos\eta_k & -\sin \eta_k & \cos\eta_k & -\sin \eta_k & 1
\end{bmatrix}}_{{\bf H}_k} \underbrace{\begin{bmatrix}
a\cos b \\
a\sin b \\
\cos b \\
\sin b \\
c \\
\end{bmatrix}}_{\bf y} + v_k \\
\end{aligned}
\endgroup
\label{step1Haupt}
\end{equation}
The choice of change of variable is arbitrary and leads to the following cost functions:
\begin{equation}
J_y =  (\bf{Z} - \mathcal{H}{\bf y})^T{\bf R}^{-1}(\bf{Z} -\mathcal{H}{\bf y})
\end{equation}
where $\mathcal{H}$ is given by:
\begin{equation}
\begingroup 
\setlength\arraycolsep{1.5pt}
\renewcommand{\arraystretch}{0.7}
\mathcal{H}  = \begin{bmatrix}
{\bf H}_1 \\
\vdots \\
{\bf H}_N
\end{bmatrix} 
\endgroup
\end{equation}
Note that even though the choice of new variable ${\bf y}$ is arbitrary, it actually dictates the condition number of $\mathcal{H}$. If $\mathcal{H}$ is not well conditioned, the result of the first stage can be poor. For this particular problem, it can be problematic if the data length $N$ is small. This is because columns 1 and 2 of ${\bf H}_k$ are same as columns 3 and 4 respectively in Eq. (\ref{step1Haupt}). This is also a pitfall of Benchmark 2.  The first-step state $\boldsymbol{y}$ is estimated using the linear least squares method.  In the second step, the estimates of the first-step states $\boldsymbol{\hat{y}}$ are treated as the new measurements in the second stage. This leads to the following measurement equation:
\begin{equation}
\hat{\bf y} = {\bf f}({\bs \xi}) + {\bf e}
\end{equation}
where the measurement noise ${\bf e}$ has covariance matrix ${\bf P}_y$. Once the estimate $\hat{\bf y}$ is obtained, the following cost function is minimized using an iterative nonlinear optimizer
\begin{equation}
J({\bs \xi}) = [\hat{\bf y} - {\bf f}({\bs \xi})]^T {\bf P}_y^{-1} [\hat{\bf y} - {\bf f}({\bs \xi})]
\end{equation}
This second-step cost function can be nonlinear and non-convex.  Thus, there is no guarantee the solution is optimal. For a static problem, this essentially reduces to solving a set of simultaneous, nonlinear algebraic equations.  In general, the solution for such problem is not unique.

\subsection{Performance Comparisons}

A set of 1000 Monte Carlo (MC) simulation runs were used to assess the performance of the algorithm developed in this paper and compare it against the two benchmarks.  For each MC run of the proposed algorithm, the initial value of the parameter $b$ is determined to be 0.08 from the first stage, and used for the second step optimization. For the first benchmark (the classic nonlinear program), initial conditions for $a$ and $c$ were selected randomly from $N\left(0,1^2\right)$ and $b$ is drawn from $\sim N\left(0,0.1^2\right)$. We also set the constraint for $b \in [0,0.2]$ in the first benchmark for a fair comparison because we only sampled  ${\bs \xi}_{2p}$ from a pre-determined range (assumed to be due to prior knowledge). The second benchmark (Haupt/Kasdin two-step estimator) does not require an initialization for the first step states, but the initial values for $a$, $b$ and $c$ are needed for the second stage. The same initial values from the 1000 runs in benchmark 1 were used for the initialization in the second stage in Benchmark 2. Benchmark 2 does not require any constraint setting for $b$, according to Ref. \cite{Haupt1996}. 
 
Table \ref{table1} shows the MC results in terms of the percentage of times the algorithm converged to the correct solution. The correction solution is determined by taking the 2-norm between the estimated and true parameter vector, that is less than 0.1.  Both Benchmark 1 and the proposed algorithm converged 100\% of the time. While this is not a theoretical proof that the correct solution is guaranteed by the algorithm developed in this paper, the comparison shows that it can yield equivalent or favorable results when compared to other nonlinear estimators. Table \ref{Ex1Result} shows the estimated parameter, standard deviation and noise covariance versus the true values. The standard deviation in the proposed estimator are calculated by taking the square root of the diagonal of the inverse of the final Hessian matrix, which is one of the outputs from \texttt{fmincon}. 

\begin{table}[ht]
	\centering
	\caption{Monte Carlo simulation results for the measurement model in Eq. (\ref{model1})}
	\begin{tabular}{cccc}
		\specialrule{.1em}{.05em}{.05em}
		\specialrule{.1em}{.05em}{.05em} 
		Estimation Algorithm  & Non-linear Programming  & Haupt/Kasdin Two-Step Estimator  & Proposed Algorithm   \\ \hline
		Correct Solution (\%)  & 100  & 97.2 & 100 \\ 
		\specialrule{.1em}{.05em}{.05em}
		\specialrule{.1em}{.05em}{.05em} 
	\end{tabular}
	Note: the estimate parameter $\hat{\xi}$ is considered correct when $\left\Vert \xi_{true} -\hat{\xi} \right\Vert_2 \leq 0.1$.
	\label{table1}
\end{table}

\begin{table}[ht]
	\caption{Estimation results from proposed \\ algorithm for scalar example 1.}
	\centering
	\begin{tabular}{cccc}
		\specialrule{.1em}{.05em}{.05em}
		\specialrule{.1em}{.05em}{.05em} 
		Parameter  &  True & Estimate & Standard Deviation \\ 
		\specialrule{.1em}{.05em}{.05em} 
		$a$  & 1 & 0.9512 & 0.0507  \\ 
		$b$  & 0.1 & 0.0812 & 0.0182  \\ 
		$c$ &  1  & 0.9415  & 0.0360 \\ 
		$R$   & 0.1237  & 0.1199 & --\\ 
		\specialrule{.1em}{.05em}{.05em}
		\specialrule{.1em}{.05em}{.05em} 
	\end{tabular} \\
	Note: The true ${\bf R}$ is calculated using the 100 noise $v$ samples. 
	\label{Ex1Result}
\end{table}

\begin{figure}[!ht]
	\centering
	\includegraphics[width=.5\textwidth]{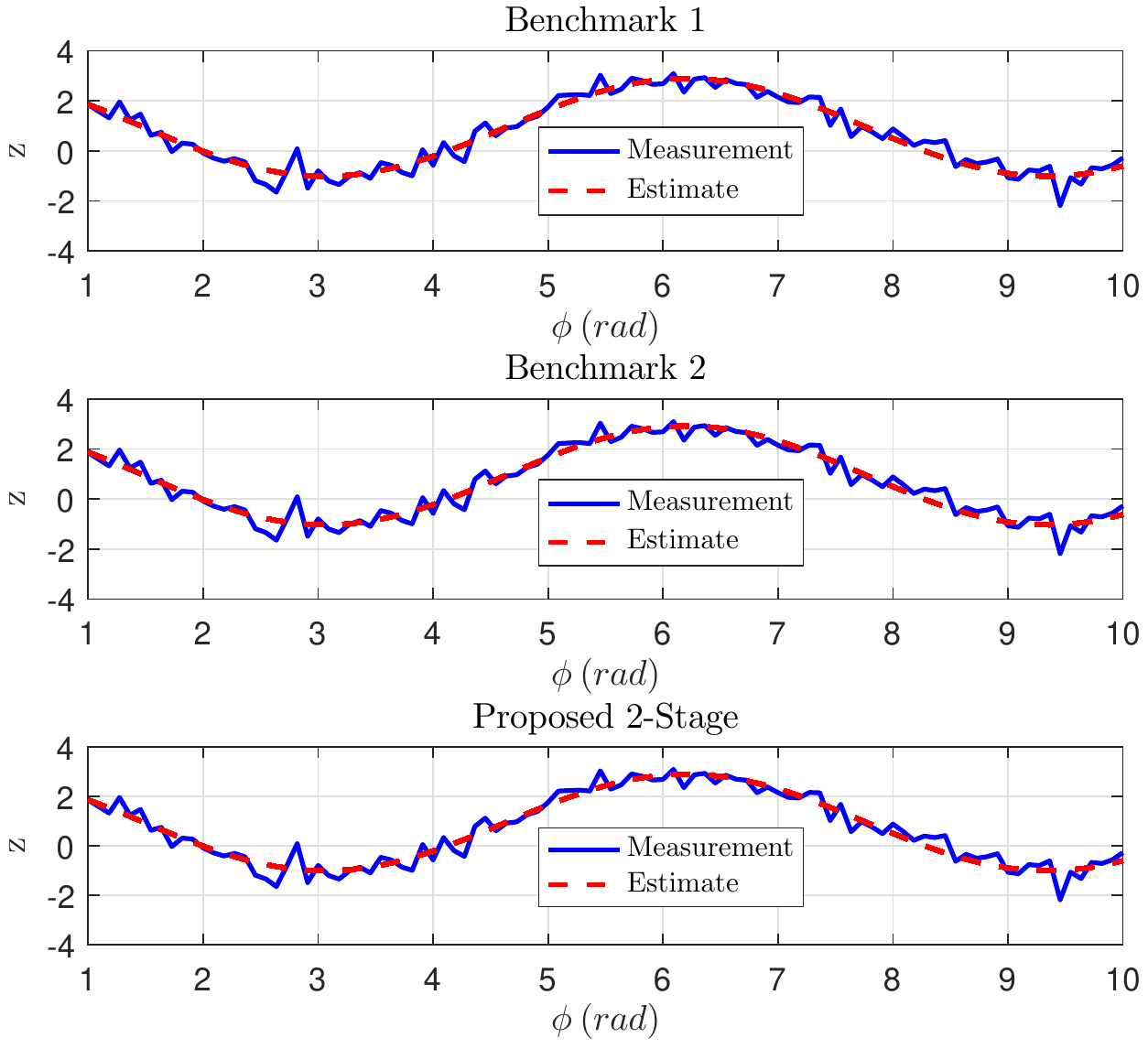}
	\caption{Random 100 MC simulation results from 3 different methods using Eq. (\ref{model1}). In this case, all 3 estimates were effective compared to the measurement.}	
	\label{methodComparison}
\end{figure}

Note that it is not always obvious (particularly, in actual applications) whether the estimator has converged to the correct solution. This can be seen if we use the estimates for the parameters to construct a predicted measurement $\hat{z}$.  That is, we apply $\hat{\bs \xi}={\begin{bmatrix} \hat{\bs \xi}_1^T & \hat{\bs \xi}_2^T \end{bmatrix}}^T $ to Eq. (\ref{model1}) to determine $\hat{z}$.  Figure \ref{methodComparison} plots 100 randomly-selected estimated outputs out of the 1000 MC runs for the proposed algorithm and the two benchmarks. In the case of the Haupt/Kasdin two-step estimator (Benchmark 2), we see that there are many instances where predicted measurement $\hat{z}$ is close to the observed measurement $z$, even though the estimates of $a$, $b$ and $c$ used to generate $\hat{z}$ are \emph{incorrect}.  The fact that the solution has converged to the incorrect value is not visible in the output.  This implies that the cost function used in the second step optimization of Haupt/Kasdin algorithm is non-convex; it has multiple local minima which are sensitive to the values of the states used to initialize the optimization process.

The comparisons so far show that breaking the estimation process into two steps can improve the chance of converging to the correct solution.  As the authors of Ref. \cite{Haupt1996} note, however, it may not always be possible to do this with the Haupt/Kasdin algorithm because of the mathematical structure of the problem at hand.  To show this, we modify the estimator problem given by Eq. (\ref{model1}) slightly as follows:

\begin{equation}
z_k = f_k(\eta_k) + v_k= (1+a)\cos(\eta_k(1+b) + c) + d + v_k
\label{model2}
\end{equation}
Equation (\ref{model2}) can be recast into the suitable form for the proposed algorithm shown below: 

\begin{equation}
\begingroup 
\setlength\arraycolsep{1.5pt}
\renewcommand{\arraystretch}{0.7}
z_k =\underbrace{\left[\begin{array}{cc}
	\cos((\eta_k + b)+c) & 1
	\end{array}\right]}_{{\bf A}\left({\bs \xi}_2\right)} \underbrace{\begin{bmatrix}
	a \\
	d \\
	\end{bmatrix}}_{{\bs \xi}_{1}}+ \underbrace{\cos((\eta_k + b)+c)}_{{\bf b}\left({\bs \xi}_2\right)} + v_k
\endgroup
\label{method3}
\end{equation}
where ${\bs \xi}_{1} = [a,d]^T$ and ${\bs \xi}_2 = [b,c]^T$. An additional unknown parameter $d$ has been added to the measurement model. In this case, the Haupt/Kasdin estimator cannot be used as the parameter $b$ cannot be linearly separated by change of variables from $\eta$. For completeness, we ran another set of Monte Carlo simulations to compare the performance of the proposed estimator and Benchmark 1 on the modified in Eq. (\ref{model2}). We draw $a$ and $d$ from $N\left(0,1^2\right)$ and $b$ and $c$ from $N\left(0,0.1^2\right)$ respectively. We also re-draw from the noise term $v_k$ from $N\left(0,0.3^2\right)$.

\begin{figure}[!ht]
	\centering
	\includegraphics[width=.5\textwidth]{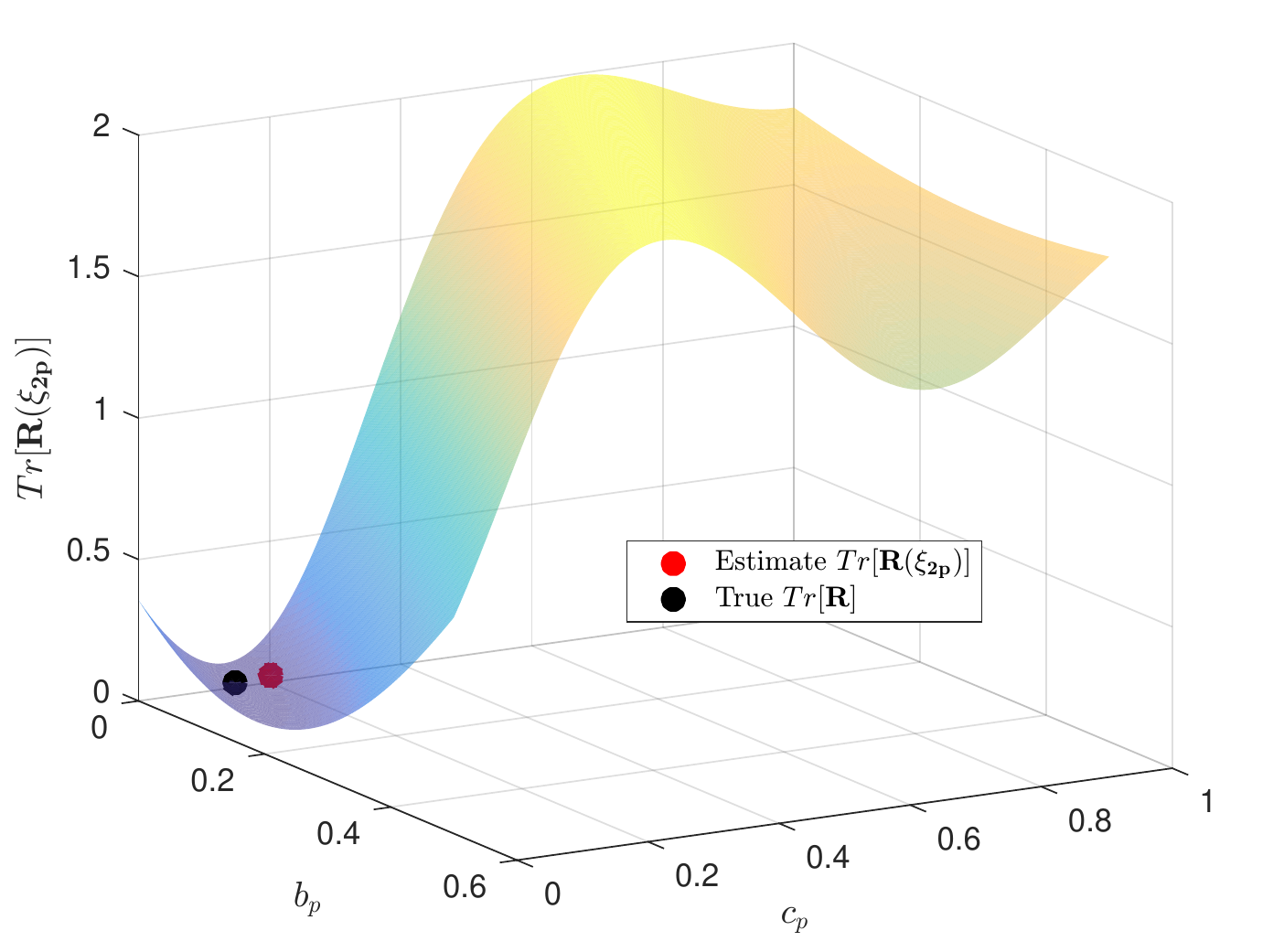}
	\caption{Estimated $\mbox{Tr} \big[{\bf R}({\bs \xi}_{2p})\big]$  by sampling random of $\xi_{2p}$ for scalar example 2.}
	\label{xi_2p_1_Ex2}
\end{figure}

\begin{figure}[!ht]
	\centering
	\includegraphics[width=.5\textwidth]{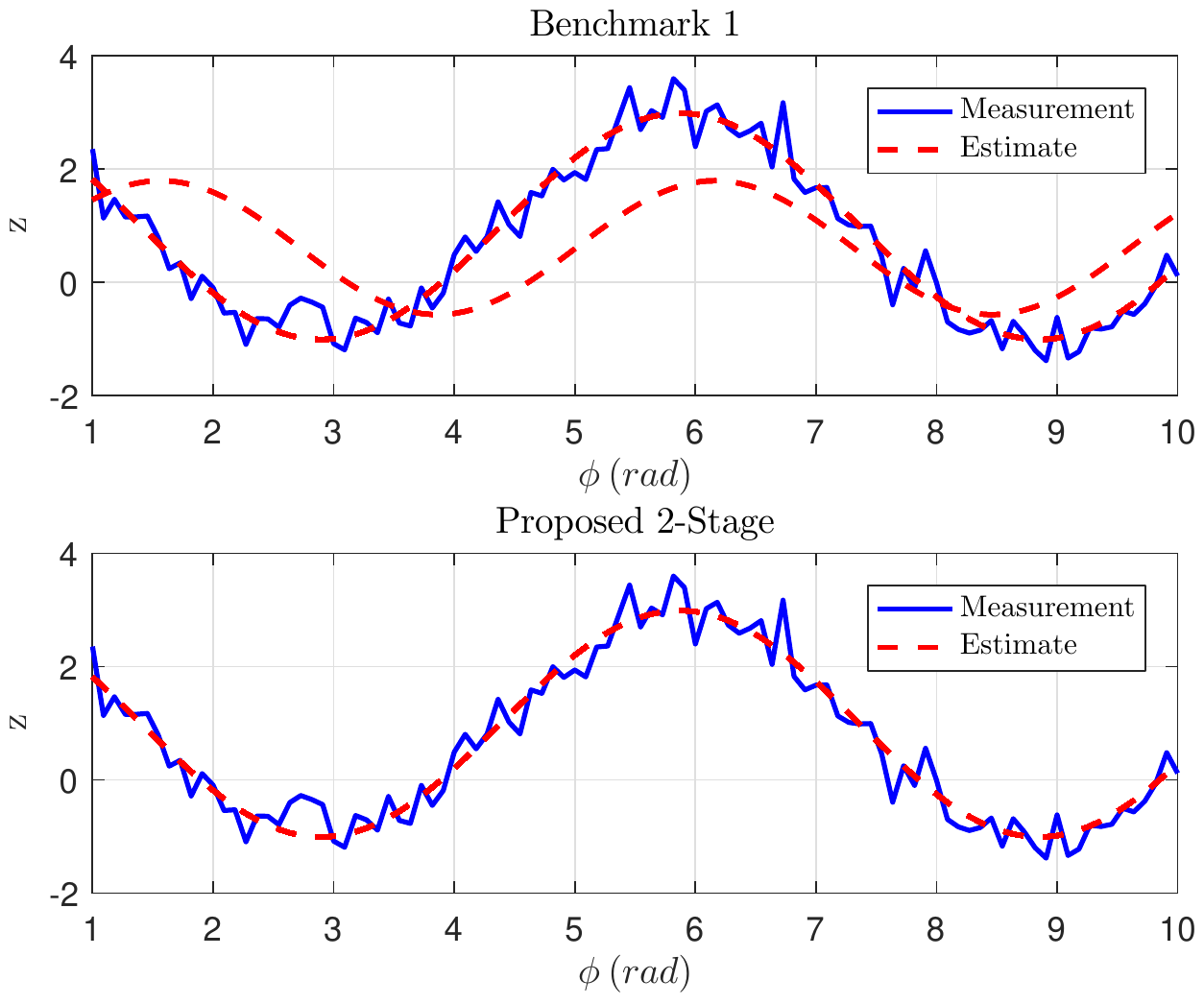}
	\caption{Random 100 MC simulation results from 2 different methods using Eq. (\ref{model2}). Benchmark 1 occasionally fails to converge to the correct estimates, whereas the proposed algorithm works consistently.}
	\label{methodComparison2}
\end{figure}

Figure \ref{xi_2p_1_Ex2} shows estimated $\mbox{Tr} \big[{\bf R}({\bs \xi}_{2p})\big]$ from sampling $b$ and $c$. It can be seen there is clearly a local minimum $\mbox{Tr} \big[{\bf R}({\bs \xi}_{2p})\big]$ value. Therefore, we use the corresponding ${\bs \xi}_{2p} = [b_p, c_p]^T = [0.0556,0.0808]^T$ as the initial condition for the second stage in the proposed estimator. We set the constraint for $b \in [0,0.5]$ and $c \in [0,1]$ in Benchmark 1 for a fair comparison because we sampled those values to estimate $\mbox{Tr} \big[{\bf R}({\bs \xi}_{2p})\big]$ for the proposed algorithm (assumed to be due to prior knowledge). The results of this simulation are summarized in Table \ref{table2}, Table \ref{Ex2Result} and Fig. \ref{methodComparison2}. It can be seen the correct percentage actually decreased due to the high nonlinearity for Benchmark 1. There are still a number of incorrect solutions, whereas the proposed algorithm still converges to the correct value every time. We randomly plotted 100 corresponding time-series of the predicted measurements out of the 1000 MC runs in Fig. \ref{methodComparison2}. It can be seen that the predicted measurement (generated by estimated parameters from the nonlinear programming approach) can be incorrect. 
    
\begin{table}[ht]
	\centering
		\caption{Monte Carlo simulation results for the measurement model in Eq. (\ref{model2}). Note that there is no entry for the Haupt/Kasdin estimator because the measurement model cannot be easily cast into a linear first-step.}
	\begin{tabular}{cccc}
		\specialrule{.1em}{.05em}{.05em}
		\specialrule{.1em}{.05em}{.05em} 
		Estimation Algorithm  & Non-linear Programming & Haupt/Kasdin Two-Step Estimator  & Proposed Algorithm   \\ \hline
		Correct Solution(\%)  & 97.8 & $-$  & 100   \\ 
		\specialrule{.1em}{.05em}{.05em}
		\specialrule{.1em}{.05em}{.05em} 
	\end{tabular}
	Note: the estimate parameter $\hat{\xi}$ is considered correct when $\left\Vert \xi_{true} -\hat{\xi} \right\Vert_2 \leq 0.1$.
	\label{table2}
\end{table}

\begin{table}[ht]
	\caption{Estimation results from proposed \\ algorithm for scalar example 2.}
	\centering
	\begin{tabular}{cccc}
		\specialrule{.1em}{.05em}{.05em}
		\specialrule{.1em}{.05em}{.05em} 
		Parameter  &  True & Estimate & Standard Deviation \\ 
		\specialrule{.1em}{.05em}{.05em} 
		$a$  & 1 & 0.9980 & 0.0474 \\ 
		$b$  & 0.05 & 0.0546 & 0.0059  \\ 
		$c$ &  0.1  & 0.0895  & 0.0351 \\ 
	    $d$ &  1  & 0.9897  & 0.0333 \\ 
		$R$   & 0.1016 & 0.1008 & --\\ 
		\specialrule{.1em}{.05em}{.05em}
		\specialrule{.1em}{.05em}{.05em} 
	\end{tabular} \\
	Note: The true ${\bf R}$ is calculated using the 100 noise $v$ samples. 
	\label{Ex2Result}
\end{table}

These two tutorial examples show that the proposed estimator can work well if the starting initial guess ${\bs \xi}_{2p}$ is close to the true value. The estimates of the first stage essentially bring the total cost very close to the true cost, which makes the nonlinear, iterative optimization of the second state converge consistently. It does this by eliminating the randomness of the initial guesses for the parameters in either two benchmark methods.

\section{Flight Test Example: 5-Hole Pitot Tube Calibration}\label{aeroEx}

Some parameter estimation problems, such as the magnetometer calibration and data compatibility problem, can be recast (shown in Appendix B) and solved with the proposed estimator. The magnetometer calibration and data compatibility problem can also be solved by well-known methods such as the Haupt/Kasdin two-step estimator and output-error, respectively. However, there are other parameter estimation problems that cannot be easily solved with these known methods because of the sensitivity to initial values. In this section, we demonstrate an aerospace application using the proposed estimator that overcomes the initial-value sensitivity issue.

In particular, we exercise the estimator on calibration of a 5-hole Pitot tube using flight test data for small UAV applications. The problem was previously investigated in Ref. \cite{Sun2019} and is an excellent example that shows how conventional methods may suffer from an incorrect local minimum, due to a poor initial parameter guess. To briefly summarize, this is the problem of calibrating a 5-hole Pitot tube (i.e., finding error model parameters) using an existing navigation solution, such as inertial velocity and attitude. The calibration consists of estimating sensor scale factor, bias errors, installation misalignment error, and steady wind vector. One challenging part of this problem is that the wind vector cannot be assumed to be zero, due to the relatively slow airspeed (10-25 m/s) range relative to the wind speed (1-10 m/s). Single-stage estimators will not converge to the correct solution if the initial parameter guess is not close to the underlining true values. In particular, the typical zero-value initial guess for wind vector might not always result in consistent estimates (i.e., the same local minimum) due to the non-zero wind vector and high nonlinearity in the measurement model.    

Since many of the details are discussed in detail in Ref. \cite{Sun2019}, we only present information required to facilitate understanding. The flight test was conducted on an Ultra Stick 120 UAV. The Ultra Stick 120 was initially used as a low-cost
flight test platform at NASA Langley Research Center \cite{Owens2006}. The Ultra Stick 120 is equipped with a traditional Pitot-static tube, a 5-hole probe \cite{Rosemount1988}, a GPS (u-blox-Neo-M8N), an Inertial Measurement Unit (IMU) (Invensense MPU-9250) and a camera. The on-board software provides a Global Navigation Satellite System/Inertial Navigation System (GNSS/INS) navigation solution in real time through an open source flight control system \cite{umnUAVlab}. 

Equations (\ref{stateinfo}a-d) show the states, input, output and parameters to be estimated. All the states in Eq. (\ref{stateinfo}a) are assumed to be known or measured from the on-board navigation solution. The input ${\bf u}$ in Eq. (\ref{stateinfo}b) are the direct pressure measurements from the 5-hole probe. The output ${\bf z}$ is the inertial velocity, resolved in the north–east–down (NED)-frame, which is also from the navigation solution. The estimated parameter ${\bs \xi}$ includes airspeed scale factor $\lambda_{V_a}$ and bias $b_{V_a}$, angle-of-attack and sideslip scale factors and biases $\lambda_{\alpha}, b_{\alpha}, \lambda_{\beta}, b_{\beta}$, installation misalignment angle $	\epsilon_{\phi}$ of the 5-hole probe rotated about the longitudinal axis of the fixed-wing aircraft, and the steady wind vector components $W_N ,W_E,W_D$. Those parameters are known to be observable through various flight excitation (wind circle, pushover-pullup, pitch chirp, yaw chirp, rudder doublet, and multisines) as described in Ref.  \cite{Sun2019a,Sun2019}. Table \ref{SamplingTime} summarizes the input design, time specifications and where this data is used in the proposed algorithm \cite{Sun2019}. Note that only those design inputs are used for the calibration - the estimated results are validated with the entire flight trajectory.

\begin{subequations}
	\begingroup 
	\setlength\arraycolsep{2.5pt}
	\renewcommand{\arraystretch}{0.7}
	\begin{align}
	{\bf x} &= \begin{bmatrix}  p & q & r &  b_{g_x} & b_{g_y}  & b_{g_z}  & \phi & \theta & \psi \end{bmatrix}^T \\
	{\bf u} &= \begin{bmatrix} P_{\Delta \alpha} &  P_{\Delta \beta} &  P_{t} & P_s &  \end{bmatrix}^T \\
	{\bf z} &= \begin{bmatrix} V_N & V_E & V_D  \end{bmatrix}^T\\ 
	{\bs \xi} &= \begin{bmatrix}
	\lambda_{V_a} &
	b_{V_a}&
	\lambda_{\alpha}  &
	b_{\alpha}  &
    \lambda_{\beta}  &
	b_{\beta} &
	\epsilon_{\phi} &
	W_N &
	W_E &
	W_D &
	\end{bmatrix}^T
	\end{align}
	\endgroup
	\label{stateinfo}
\end{subequations}

\begin{table}[tbh]
	\caption{Input design and time specification for calibration.}
	\centering
	\scalebox{1}{
		\begin{tabular}{ccc}
		\specialrule{.1em}{.05em}{.05em}
		\specialrule{.1em}{.05em}{.05em} 
			Maneuver Type &  Time (sec) & Usage  \\ 
			\hline
			Wind Circle 1 &  [384, 408.2] & Stage 1    \\
			Wind Circle 2 &  [411, 438.3] & Stage 1   \\  
			Pushover-pullup (POPU) &  [510.9, 530]  & Stage 1  \\ 
			Multisine 1 &  [576, 596] & Stage 2  \\
			Multisine 2 &  [690, 711] & Stage 2  \\
			Multisine 3 &   [752, 772] & Stage 2  \\
			Multisine 4 &   [810, 830] & Stage 2  \\
			Pitch Chirp 1     &  [867, 887]   & Stage 2  \\
			Yaw Chirp     &  [980.080, 1013.595]  & Stage 2  \\	
			Pitch Chirp 2     &  [1041, 1061]   & Stage 2  \\
			Rudder Doublet &  [1113, 1115]   & Stage 2  \\		
		\specialrule{.1em}{.05em}{.05em}
		\specialrule{.1em}{.05em}{.05em} 
		\end{tabular}
	}
	\label{SamplingTime}
\end{table}

Equation (\ref{errorEq}) shows the air data error model. Though linear in the unknown parameters, it is determined to be sufficient for capturing the error dynamics in this 5-hole probe (Sec. II in \cite{Sun2019}). 

\begin{equation}
\begingroup 
\setlength\arraycolsep{1.0pt}
\renewcommand{\arraystretch}{0.7}
\begin{aligned}
V_a &= (1+\lambda_{V_a})\sqrt{\frac{2(P_t - P_s)}{\rho}} + b_{V_a}\\
\alpha &= (1+\lambda_{\alpha}) \frac{P_{\Delta \alpha} }{K_{\alpha} (P_t - P_s)} + b_{\alpha}\\
\beta &= (1+\lambda_{\beta}) \frac{P_{\Delta \beta} }{K_{\beta} (P_t - P_s)} + b_{\beta}\\
\end{aligned}
\endgroup
\label{errorEq}
\end{equation}

Equation (\ref{windTri}) is the wind triangle equation resolved in the NED frame. The vector $\mathbf{V}$ and $\mathbf{W}$ are inertial vector and wind vector resolved in the NED frame. The vector $\mathbf{V}_{a,cg}$ is the airspeed vector (which consists of the body-axis translational components) at the center of IMU (in this case very close to the center of gravity, hence denoted with the subscript $cg$) and $\mathbf{V}_{a,s}$ is the airspeed vector at the 5-hole probe sensor location. The matrix $C^n_b$ is the coordinate transformation from body frame to inertial frame and $C(\epsilon_{\phi})$ accounts for installation misalignment angle $\phi$ rotated about the longitudinal axis. Finally, $\mathbf{\omega}$ and ${\bf r}$ are the corrected rotational velocity and displacement vector from the center of the IMU in the UAV to the 5-hole probe sensor location. The exact formulation of 
$C^n_b$, $C(\epsilon_{\phi})$, $\mathbf{\omega}$ and ${\bf r}$ are shown in Eq. (\ref{wingTri2}) and (\ref{Cbn}).

\begin{equation}
\mathbf{V}
= C^n_b \mathbf{V}_{a,cg} + \mathbf{W} = C^n_b  \big[C(\epsilon_{\phi}) \mathbf{V}_{a,s} -  [\mathbf{\omega}]_{\times} \mathbf{r}\big] + \mathbf{W}
\label{windTri}
\end{equation}

\begin{equation}
\begingroup 
\setlength\arraycolsep{1.0pt}
\renewcommand{\arraystretch}{0.7}
\mathbf{\omega} = \begin{bmatrix}
p - b_{p} \\
q - b_{q} \\
r - b_{r} \\
\end{bmatrix} \hspace{0.2in}
\mathbf{r} = \begin{bmatrix}
x_s \\
y_s \\
z_s \\
\end{bmatrix}
\hspace{0.2in}
C(\epsilon_{\phi}) = \begin{bmatrix}
1 & 0 & 0 \\
0 & cos\epsilon_{\phi}   & sin\epsilon_{\phi}  \\
0 & -sin\epsilon_{\phi} & cos\epsilon_{\phi} 
\end{bmatrix}
\endgroup
\label{wingTri2}
\end{equation}

\begin{equation}
\begingroup 
\setlength\arraycolsep{1.0pt}
\renewcommand{\arraystretch}{0.7}
C^n_b = \begin{bmatrix}
cos\theta cos\psi & sin\phi sin\theta sin\psi - cos\phi sin\psi & cos\phi sin\theta cos\psi + sin\phi sin\psi \\
cos\theta sin\psi & sin\phi sin \theta sin \psi +cos\phi cos\psi &
cos\phi sin\theta \sin \psi - sin\phi cos\psi \\
-sin\theta & sin\phi cos \theta & cos \phi cos \theta  
\end{bmatrix}
\endgroup
\label{Cbn}
\end{equation}

We use the wind triangle equation in Eq. (\ref{windTri}) as the measurement equation with assumed additive Gaussian white noise $[v_{V_N} \: v_{V_E} \: v_{V_D}]^T$ to represent measurement noise and to recast it into a suitable form for the two-stage estimator as follows:  

\begin{equation}
\begingroup 
\setlength\arraycolsep{3.0pt}
\renewcommand{\arraystretch}{0.7}
\begin{aligned}
{\bf z}_k = \begin{bmatrix}
V_N \\
V_E \\
V_D \\
\end{bmatrix}_k  &= \underbrace{\begin{bmatrix}
	 F\sqrt{\frac{2(P_t - P_s)}{\rho} }  & F  & I_3 
	\end{bmatrix}}_{{\bf A}({\bf x},{\bf u},{\bs \xi}_2)} \underbrace{\begin{bmatrix}
	\lambda_{V_a} \\
	b_{V_a}  \\
	W_N \\
	W_E \\
	W_D \\
	\end{bmatrix}}_{{\bs \xi}_{1}} + \underbrace{ F \sqrt{\frac{2(P_t - P_s)}{\rho} }   - C^n_b  [\mathbf{\omega}]_{\times} \mathbf{r}
    }_{{\bf b}({\bf x}, {\bf u},{\bs \xi}_2)}  + \underbrace{\begin{bmatrix}
	v_{V_N} \\
	v_{V_E}  \\
	v_{V_D}  \\
	\end{bmatrix}_k}_{{\bf v}_k}
\end{aligned}
\endgroup
\label{airdataModelLon}
\end{equation}
where the $F$ is a 3 by 1 vector 
\begin{equation}
\begingroup 
\setlength\arraycolsep{0.7pt}
\renewcommand{\arraystretch}{0.7}
F = 
 C^n_b C(\epsilon_{\phi})\begin{bmatrix}
	cos\alpha cos \beta \\
	sin \beta  \\
	sin\alpha cos \beta \\
\end{bmatrix}
\endgroup
\end{equation}
The parameter vector ${\bs \xi}$ is now separated into ${\bs \xi}_1$ and ${\bs \xi}_2$ as shown in Eq. (\ref{parameter2}). 

\begin{subequations}
	\begingroup 
	\setlength\arraycolsep{2.5pt}
	\renewcommand{\arraystretch}{0.7}
	\begin{align}
	{\bs \xi}_1 &= \begin{bmatrix}
	\lambda_{V_a} &
	b_{V_a}&
	W_N &
	W_E &
	W_D &
	\end{bmatrix}^T\\ 
	{\bs \xi}_2 &= \begin{bmatrix}
	\lambda_{\alpha}  &
	b_{\alpha}  &
	\lambda_{\beta}  &
	b_{\beta} &
	\epsilon_{\phi} &
	\end{bmatrix}^T
	\end{align}
	\endgroup
\label{parameter2}
\end{subequations}

We estimate the parameter vector ${\bs \xi}$ using the proposed estimator. Intuitively, the proposed estimator works for this initial-condition-sensitive calibration problem because it isolates some of the nonzero parameters (e.g., wind vector) and minimizes the cost in the first stage until it is close to the optimal cost.

\begin{figure}[!ht]
	\centering
	\includegraphics[width=.5\textwidth]{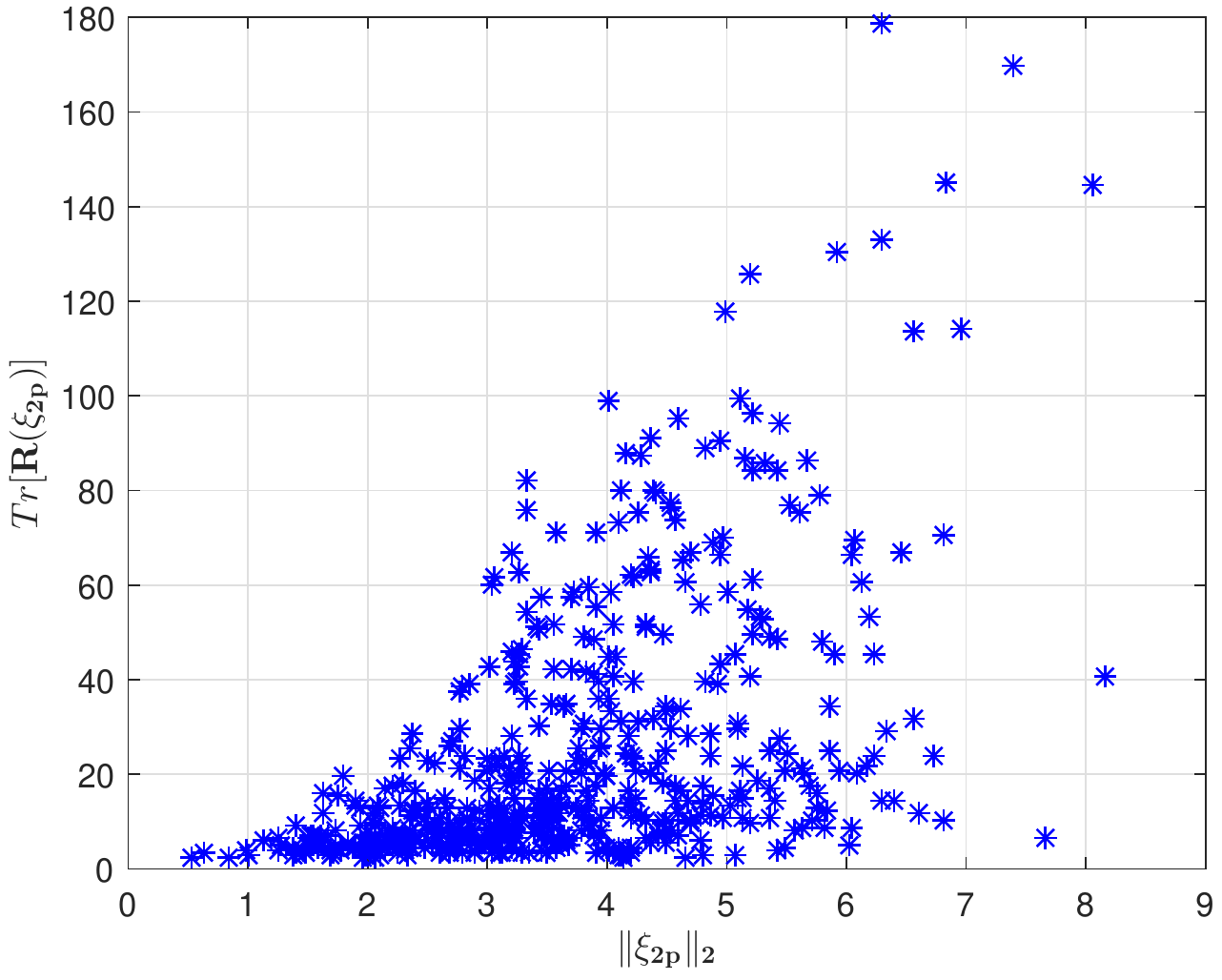}
	\caption{Estimated $\mbox{Tr} \big[{\bf R}({\bs \xi}_{2p})\big]$ versus its corresponding 2-norm of $\xi_{2p}$ using 500 random $\xi_{2p}$.}
	\label{random_xi_2p}
\end{figure}

Since the parameter ${\bs \xi}_2$ is expected to be small, or at least bounded, we sampled 500 random $\lambda_{\alpha}, b_{\alpha}, \lambda_{\beta},b_{\beta} $ from $N\left(0,2^2\right)$ and $\epsilon_{\phi}$ from $N\left(0,0.3491^2\right)$ (standard deviation of 20 deg) respectively. Figure \ref{random_xi_2p} shows $\mbox{Tr} \big[{\bf R}({\bs \xi}_{2p})\big]$ versus the 2-norm of ${\bs \xi}_{2p}$ using the 500 samples of ${\bs \xi}_2$. Notice that there is no unique local minimum (a flat region when $||{\bs \xi}_{2p}||_2 = 1$ to $4$) using the sampled values, which means the second order condition is close to zero. This also means that the nonlinear estimator with respect to ${\bs \xi}_2$ only might not work well since the inner optimization in Eq. (\ref{equalivant}) may not have brought the estimated cost close enough to the true cost, so ${\bs \xi}_1$ cannot be uniquely determined in the minimal residual sense. Hence, we have to re-estimate ${\bs \xi}_1$ and ${\bs \xi}_2$ simultaneously with an initial guess ${\bs \xi}_{2p}$ and the estimated $\hat{\bs \xi}_{1p}$ from the first stage. The initial guess of  ${\bs \xi}_{2p} = [0.2, -1 ,-0.2, -1 , -0.1]^T$ was determined to be a good initial condition from the 500 samples. Note the 2-norm of ${\bs \xi}_{2p}$ should still be small based on Fig. \ref{random_xi_2p}.   

Table \ref{table3} shows the final estimated parameters and the associated standard deviations in parentheses. It also lists the constraints used in the second stage, which was determined by the physical limitations of the system. The constraints of the wind vector were also refined based on the output of the first stage.

\begin{table}[ht]
	\caption{Parameter estimate, standard deviation and constraint setting.}
	\centering
	\begin{tabular}{cccc}
		\specialrule{.1em}{.05em}{.05em}
		\specialrule{.1em}{.05em}{.05em} 
		Parameter  &  Two-stage (standard deviation) & unit & Constraints used \\ 
		\specialrule{.1em}{.05em}{.05em} 
		$\lambda_{V_a}$  & -0.1748  (0.0739) & -- & [-0.5, 0.5] \\ 
		$b_{V_a}$  & 4.3553 (1.2672)  & $m/s$ & [-5, 5] \\  
		$\lambda_{\alpha}$ & 0.2982 (0.3228) & -- & [-0.5, 0.5] \\ 
		$b_{\alpha}$   & -2.4854 (0.5603)  & $deg$ & [-5, 5] \\ 
		$\lambda_{\beta}$ & -0.2673  (0.2379)   & -- & [-0.5, 0.5] \\ 
		$b_{\beta}$  &  -1.2980  (0.5164)   & $deg$ & [-5, 5] \\ 
		$\epsilon_{\phi}$  & -0.1380 (0.2699)  & $rad$ & [-0.2618, 0.2618] \\  
		$W_N$ & -3.8038 (0.8704) & $m/s$ & [-6, 6] \\ 
		$W_E$ & -2.4137 (0.9956) & $m/s$ & [-6, 6] \\ 
		$W_D$ & -0.7168 (0.9952) & $m/s$ & [-2, 2] \\ 
		\specialrule{.1em}{.05em}{.05em}
		\specialrule{.1em}{.05em}{.05em} 
	\end{tabular}
	\label{table3}
\end{table}

\begin{figure}[!ht]
	\hfill
	\subfigure[Comparison between measured and corrected inertial velocity components.]
	{
		\centering
		\includegraphics[width=.455\textwidth]{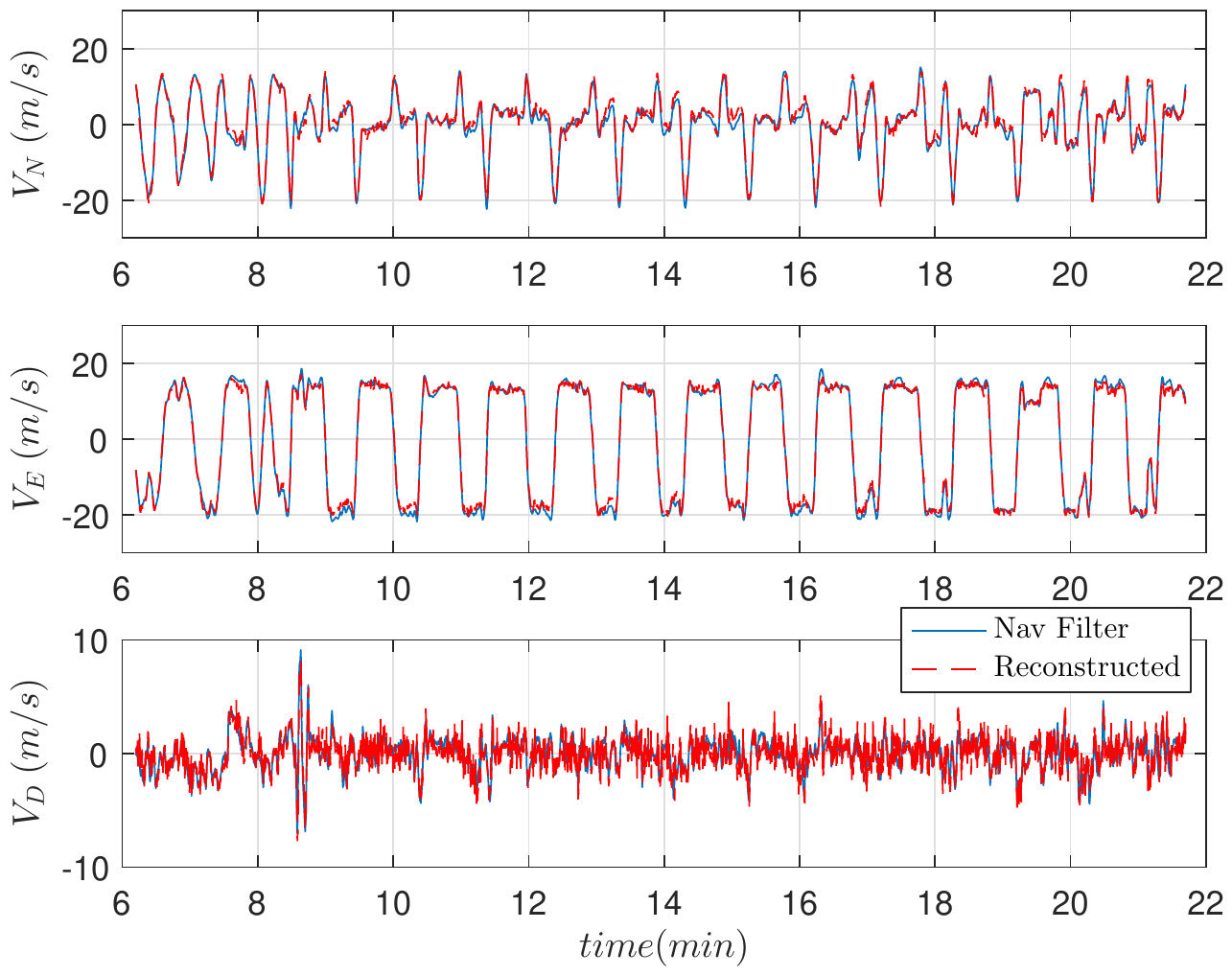}
		\label{NEDVelocityCompare}
	}
	\hfill
	\subfigure[Error between measured and corrected inertial velocity components.]
	{
		\includegraphics[width=.45\textwidth]{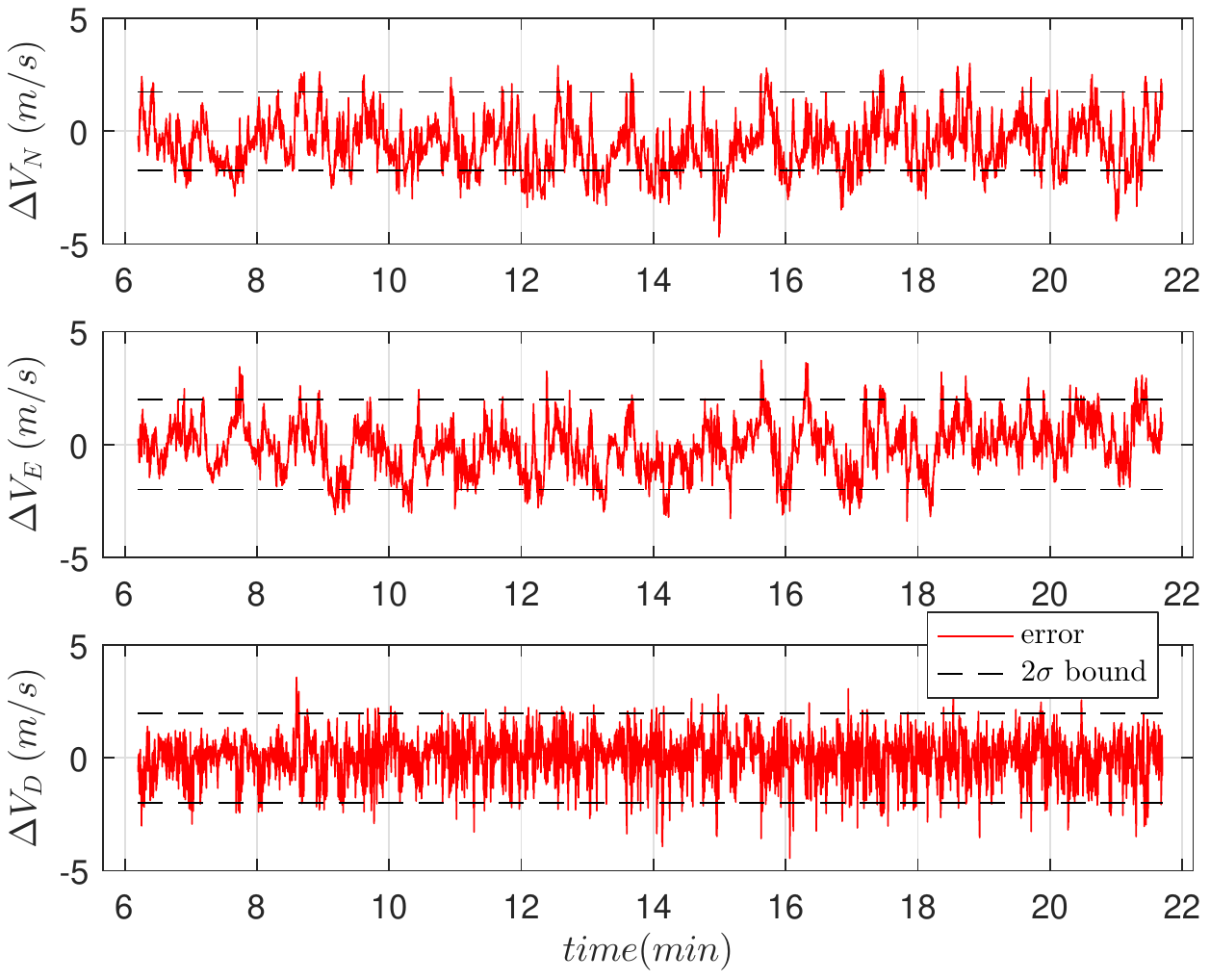}
		\label{NEDVelocityError}
	}
	\hfill
	\caption{Measured vs. corrected inertial velocity.}
	\label{NEDVelocity}
\end{figure}

Figure \ref{NEDVelocityCompare} and \ref{NEDVelocityError} show the reconstructed (estimated) and measured (GNSS/INS solution) inertial velocity components and their error over the entire flight trajectory. Table \ref{table4} lists the root mean square error values of the estimated outputs and noise standard deviation from the estimated ${\bf R}$. The estimated output matches well with the measurement; the error plot is mostly bounded by the estimated 2 standard deviations. When using the single-stage method (Benchmark 1 - not shown), there is a large discrepancy between the reconstructed and measured inertial velocity, though the estimator was able to converge. This means that without good initial guess, the single-stage estimator may not always converge to the correct minimum.  Estimating ${\bs \xi}_{2}$ only in the second stage also did not work well in terms of the error between measured and computed outputs.

\begin{table}[ht]
	\caption{Output root mean square error \\
		and measurement noise stand deviation ($\sqrt{\mbox{diag}(R)}$).}
	\centering
	\begin{tabular}{ccc}
		\specialrule{.1em}{.05em}{.05em}
		\specialrule{.1em}{.05em}{.05em} 
		&  Mean Square Error (m/s) & Estimated standard deviation (m/s)  \\ 
		\specialrule{.1em}{.05em}{.05em} 
		$V_N$ & 1.2564 &  0.9875 \\
		$V_E$ &  1.1028 & 0.8645\\
		$V_D$ &  0.8321 &  0.9207\\
		\specialrule{.1em}{.05em}{.05em}
		\specialrule{.1em}{.05em}{.05em} 
	\end{tabular}
	\label{table4}
\end{table}

\begin{figure}[!ht]
	\hfill
	\subfigure[Comparison between calibrated Pitot tube measurement and cablirated 5-hole probe estimate.]
	{
		\centering
		\includegraphics[width=.465\textwidth]{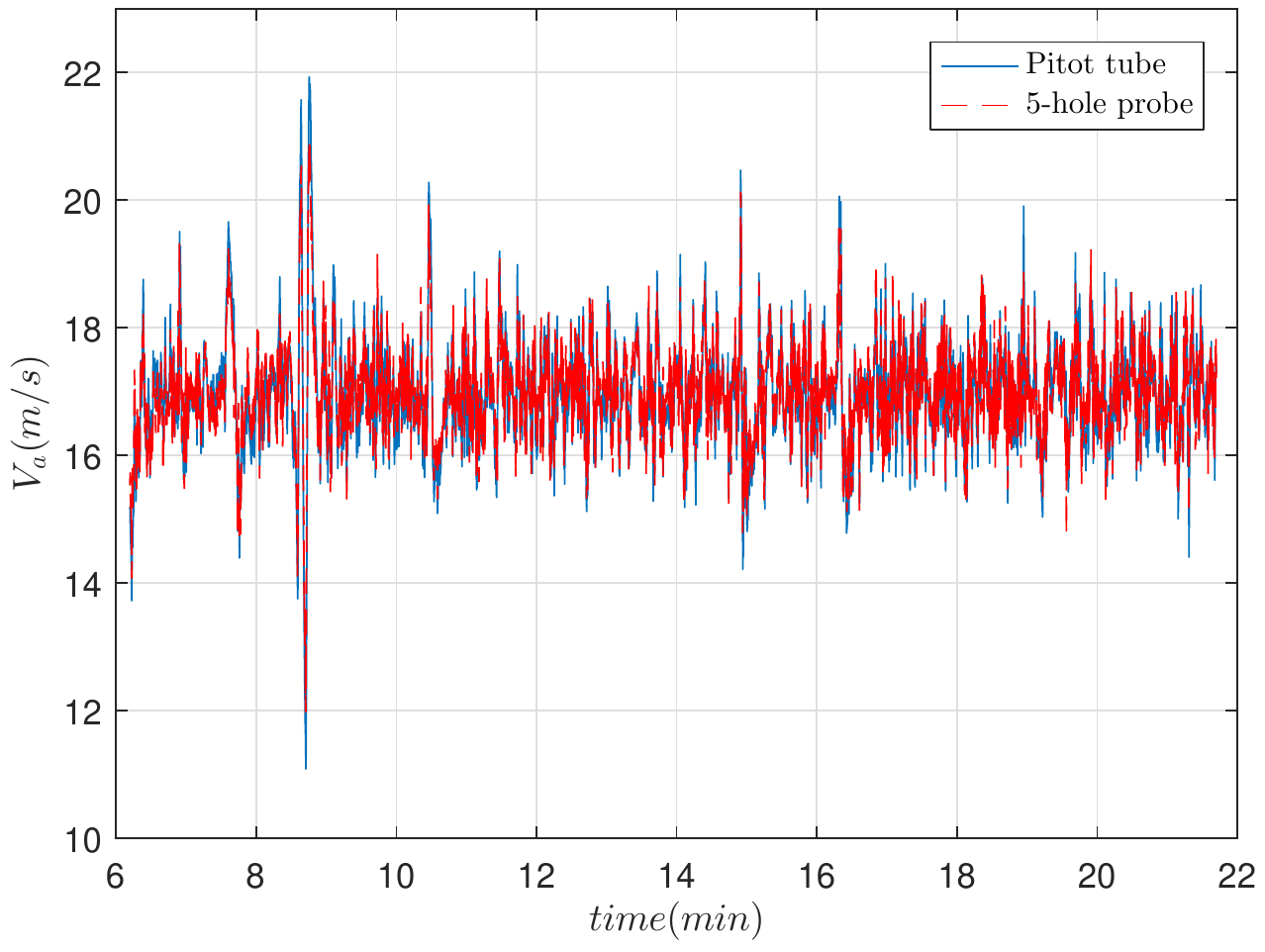}
		\label{AirspeedEst}
	}
	\hfill
	\subfigure[Error calibrated Pitot tube measurement and cablirated 5-hole probe estimate.]
	{
		\includegraphics[width=.44\textwidth]{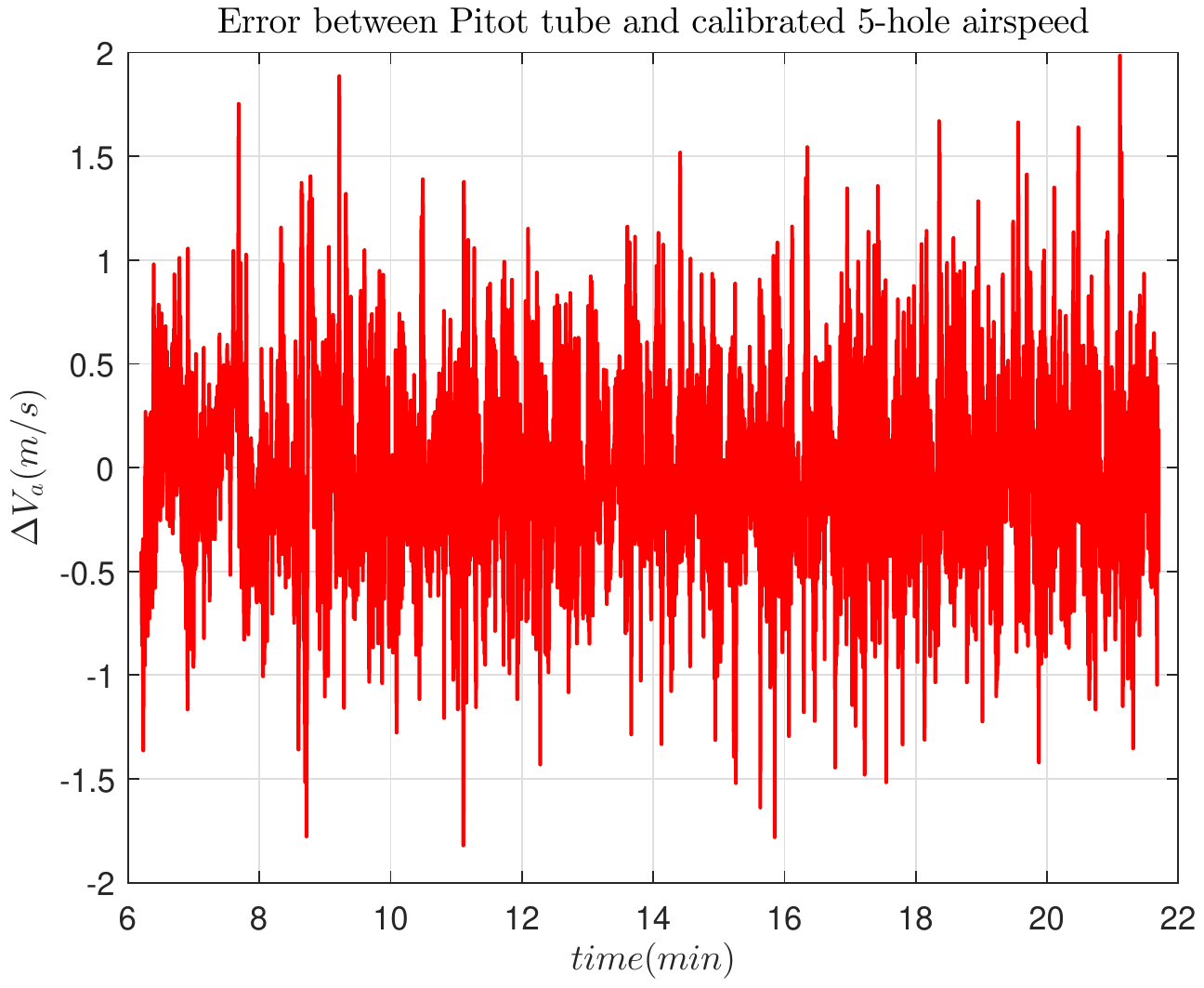}
		\label{errorVa}
	}
	\hfill
	\caption{Measured vs. estimated 5-hole airspeed measurement.}
	\label{airspeed}
\end{figure}

Figure \ref{AirspeedEst} shows the estimated airspeed from the 5-hole probe and the onboard airspeed measurement from an independently-calibrated Pitot tube. The error between the estimated and measured airspeed is shown in Fig. \ref{errorVa} and the root mean square error was calculated to be 0.1241 $m/s$.  
The small error in airspeed when compared to another independent source also supports our claim that the proposed estimator worked well for this calibration problem. 

\section{Conclusion}\label{Conclusion} 

This paper presented a two-stage estimation algorithm for solving a class of nonlinear, parameter estimation problems that appear in aerospace engineering applications. This class of problems appears as a result of the mathematical form of the standard sensor error model used.  Problems having this form can be recast into a problem that is linear with respect to a subset of the unknown parameters and nonlinear with respect to the remaining parameters. Implementation of the proposed estimator proceeds in two stages.  In the first stage, linear least squares is used to obtain initial values for a subset of the unknown parameters while a residual sampling procedure is used for selecting initial values for the rest of the parameters. In the second stage, only a subset of the parameters needs to be re-estimated, and the rest of the parameters can be immediately calculated via weighted least squares. However, if we cannot determine a unique local minimum condition for the second stage, all the parameters have to be re-estimated simultaneously by a nonlinear constrained optimization. The examples provided in this paper show that this approach alleviates the initial condition sensitivity issue and minimizes the likelihood of converging to an incorrect local minimum of the nonlinear cost function. It also provides a technique for selecting initial conditions for a nonlinear measurement model that has the same canonical form. Furthermore, it was shown that if the measurement model and unknown parameters satisfy certain conditions (i.e., Lipschitz continuity and finite domain), then the error in the final cost of the optimization has an upper bound.  

While the problems presented in this paper had static parameters, the algorithm can be used to find initial conditions with a mini-batch data set for dynamic problems as well. Therefore, we believe this algorithm is yet one more tool available to the designer of estimators for nonlinear engineering problems.

\section*{Acknowledgments}

The authors gratefully acknowledge Professor Peter Seiler of the Department of Aerospace Engineering \& Mechanics at the University of Minnesota, Twin Cities for his insightful critique, technical discussions, and comments in support of developing and validating the work presented in this paper.  The authors also gratefully acknowledge the Minnesota Invasive Terrestrial Plants and Pests Center (MITPPC) for financial support to conduct research associated with increasing the reliability of small UAV technology used for surveying applications.

\appendix	
\section*{Appendix A: Derivation of Canonical Form}\label{classficiation}
\setcounter{equation}{0}
\renewcommand{\theequation}{A\arabic{equation}}
The purpose of this appendix is to show how the model structure given by Eq. (\ref{eq:canonical_structure}) arises in aerospace estimation problems.  The structure arises from what we refer to in this paper as the \emph{standard sensor error models}.  While not referred to as such, its mathematical form is given in Ref. \cite[Eq. (10.13)]{Klein2006} and Ref. \cite[Eq (4.16) and (4.17)]{Groves} and it relates a vector measurement ${\bf z}\in \mathbb{R}_{3 \times 1}$ made by a sensor (e.g., an accelerometer triad) to the actual physical quantity being measured denoted by ${\bf y} \in \mathbb{R}_{3 \times 1}$.  Mathematically, it is the affine map from ${\bf y}$ to ${\bf z}$ given by:
\begin{align}
	{\bf z} = {\bf C}\,{\bf y} + {\bf n} + {\bf v}
\end{align}
where the entries of ${\bf C} \in \mathbb{R}_{3 \times 3}$ represent systematic errors such as scale factor deviations and axes misalignments.  The vector ${\bf n}\in \mathbb{R}_{3 \times 1}$ represents null-shifts (biases) and ${\bf v} \in \mathbb{R}^{3 \times 1}$ represents random, output noise normally modeled as a normal distribution with some given covariance.  The entries of the matrix ${\bf C}$ and ${\bf n}$ are usually unknown parameters and need to be estimated.  Discussion of the nature of the entries in ${\bf C}$, ${\bf n}$ and ${\bf v}$ is beyond the scope of this paper, but we refer the interested reader to the text by Ref. \cite[Chapter 4]{Groves} for more details.  In this appendix, we are interested in the mathematical structure of ${\bf C}$ which is normally the product of multiple matrices, each representing a different type of error.

Let us consider a typical  simple case where ${\bf C}$ is the product of two matrices:  A misalignment error matrix ${\bf C}_{\bs \eta}$ and scale factor error matrix ${\bf C}_{\bs \lambda}$.  The subscript ${\bs \eta}$ represents the vector $\renewcommand{\arraystretch}{0.7} {\bs \eta} = \small \left[\begin{array}{ccc} \eta_1 & \eta_2 & \eta_3 \end{array}\small\right]^T$ whose entries are small misalignment errors between the triads ${\bf z}$ and ${\bf y}$.  Since the entries in ${\bs \eta}$ are normally very small (i.e., $\eta_i \ll 1, \; i = 1, 2, 3$), the matrix ${\bf C}_{\eta}$ is approximated as a skew symmetric matrix of the vector ${\bs \eta}$.  Similarly, the subscript $\lambda$ represents the vector of scale factor errors $\renewcommand{\arraystretch}{0.7}  {\bs \lambda} = \small \left[\begin{array}{ccc} \lambda_1 & \lambda_2 & \lambda_3 \end{array}\small\right]^T$. The scale factor errors $\lambda_i \ll 1, \; i = 1,2, 3$ and appear on the diagonal of ${\bf C}_{\lambda}$.  This leads to ${\bf C}$ having the following structure:
\begin{align}
	{\bf C} = {\bf C}_{\bs \eta}{\bf C}_{\bs \lambda} = \begin{bmatrix}
		1 & -\eta_3 & \eta_2 \\ \eta_3 & 1 & -\eta_1 \\ -\eta_2 & \eta_3 & 1
	\end{bmatrix} \begin{bmatrix}
		1 + \lambda_1 & 0 & 0 \\ 0 & 1+\lambda_2 & 0 \\ 0 & 0 & 1 + \lambda_3
	\end{bmatrix} = \begin{bmatrix}
		1+\lambda_1 & -\eta_3 & \eta_2 \\ \eta_3 & 1 + \lambda_2 & -\eta_1 \\ -\eta_2 & \eta_3 & 1 + \lambda_3
	\end{bmatrix}
\end{align}
where we have assumed $\eta_i \eta_j = \lambda_i \lambda_j = \lambda_i \eta_j = 0$ for $i = 1, 2, 3$. Note that if ${\bs \eta}$ is not small, it still can be recast into this structure.  

This structure of the sensor output error affine map can be generalized if we replace ${\bf y}$ by ${\bf f}({\bf x}, {\bf u},\boldsymbol{\xi^{'}})$ (so that it can include the known state ${\bf x}$ as well as the unknown parameters ${\bs \xi}$ and control inputs ${\bf u}$) and write it as:
\begin{equation}
\begingroup 
\setlength\arraycolsep{1.5pt}
\renewcommand{\arraystretch}{0.7}
\begin{aligned}
{\bf z} &= \left[\prod_{m=1}^{\infty}{\bf N}_m {\bf D}\right]{\bf f}({\bf x}, {\bf u},\boldsymbol{\xi^{'}}) + {\bf n} + {\bf v}
\end{aligned}
\endgroup
\label{canonical}
\end{equation} 
where ${\bf N}_m \in \mathbb{R}_{3 \times 3}$ are \emph{non-diagonal} matrices and ${\bf D} \in \mathbb{R}_{3 \times 3}$ is a diagonal matrix. The product $\prod_{m=1}^{\infty}$ means that there can be infinitely many ${\bf N}$ matrices. In real applications, usually $m < 4$.  The function ${\bf f}$ still can have unknown parameters associated with the input ${\bf u}$, but the number of unknowns in ${\bf f}$ is reduced due to factorization of the matrices ${\bf N}_m$. We denote the reduced parameter vector as $\boldsymbol{\xi^{'}}$.

Since the unknown parameters of ${\bf D}$ are on the diagonal and the unknown bias vector ${\bf n}$ is additive, this can be transformed into the following linear affine form: 
\begin{equation}
\begingroup 
\setlength\arraycolsep{1.5pt}
\renewcommand{\arraystretch}{0.7}
\begin{aligned}
{\bf z}_{3 \times 1} &= \underbrace{\left[ \begin{array}{c|c} 
	\prod_{m=1}^{\infty}{\bf N}_m \mathcal{D}\big[{\bf f}({\bf x},{\bf u},\boldsymbol{\xi^{'}})\big] & I_{3 \times 3}
	\end{array}\right]}_{{\bf A}(\boldsymbol{\xi_2})}\underbrace{\begin{bmatrix}
	{\bf D}(1,1) \\
	{\bf D}(2,2) \\
	{\bf D}(3,3) \\
	{\bf n} \\
	\end{bmatrix}}_{\boldsymbol{\xi_1}} 
+  \underbrace{\prod_{m=1}^{\infty} {\bf N}_m {\bf f}_{3 \times 1}({\bf x}, {\bf u})}_{{\bf b}(\boldsymbol{\xi_2})} + {\bf v} 
\end{aligned}
\endgroup
\label{canonical2}
\end{equation} 
where ${\bf A}(\boldsymbol{\xi_2})$ and ${\bf b}(\boldsymbol{\xi_2})$ contain all the parameters in ${\bf N}$ and ${\bf f}$, and $\boldsymbol{\xi_1}$ represents the rest of the unknown parameters. The operator $\mathcal{D}\big[ \cdot \big]$ takes in a vector and returns a square matrix with elements of the vector on the diagonal. If there are more measurement vectors that have the same structure shown in Eq. (\ref{canonical2}), they can be concatenated as follows:
\begin{equation}
\begingroup 
\setlength\arraycolsep{1.5pt}
\renewcommand{\arraystretch}{0.7}
\begin{aligned}
{\bf Z}_{3n \times 1}
=  \begin{bmatrix}
{\bf z}_{1} \\
{\bf z}_{2} \\
\vdots \\
{\bf z}_{n} \\
\end{bmatrix} & = \begin{bmatrix}
{\bf A}_1\left({\bs \xi}_{2,1}\right) & \boldsymbol{0}_{3 \times 6} & \cdots & \cdots  & \boldsymbol{0}_{3 \times 6} &  \boldsymbol{0}_{3 \times 6} \\
\boldsymbol{0}_{3 \times 6} & {\bf A}_2\left({\bs \xi}_{2,2}\right) & \boldsymbol{0}_{3 \times 6} & \cdots & \boldsymbol{0}_{3 \times 6} &  \boldsymbol{0}_{3 \times 6}  \\
\vdots & \vdots   & \vdots    &  \vdots  &  \vdots & \vdots \\
\boldsymbol{0}_{3 \times 6}& \boldsymbol{0}_{3 \times 6} & \cdots &  \cdots& \cdots &  {\bf A}_n\left({\bs \xi}_{2,n}\right)  \\
\end{bmatrix} \begin{bmatrix}
{\bs \xi}_{1,1}\\
{\bs \xi}_{1,2} \\
\vdots \\
{\bs \xi}_{1,n} \\
\end{bmatrix} + \begin{bmatrix}
{\bf b}_1\left({\bs \xi}_{2,1}\right) \\
{\bf b}_2\left({\bs \xi}_{2,2}\right) \\
\vdots \\
{\bf b}_n\left({\bs \xi}_{2,n}\right) \\
\end{bmatrix}+ \begin{bmatrix}
{\bf v}_1 \\
{\bf v}_2 \\
\vdots \\
{\bf v}_n \\
\end{bmatrix}\\
&= {\bf A}\left({\bs \xi}_2\right){\bs \xi}_{1} + {\bf b}({\bs \xi}_2) + \boldsymbol{V} \\
\end{aligned}
\endgroup
\label{canonical3}
\end{equation} 
where ${\bs \xi}_{1} = [{\bs \xi}_{1,1}, {\bs \xi}_{1,2},\ldots,{\bs \xi}_{1,n}]^T$ and ${\bs \xi}_2 = [{\bs \xi}_{2,1}, {\bs \xi}_{2,2},\ldots,{\bs \xi}_{2,n}]^T$. The combination of  ${\bs \xi}_{1}$ and ${\bs \xi}_2$ represent the total unknown parameters vector ${\bs \xi}$. Even though the total measurement vector ${\bf Z}$ in Eq. (\ref{canonical3}) has $3n$ number of elements, it does not have to be multiple of three, depending on the given measurement model. For example, quaternion-related measurements can have  an even number of measurement equations. 

There are many parameter estimation problems that can be recast into this canonical form in the field of aerospace engineering. For example, magnetometers are used extensively in navigation, guidance and control applications \cite{Crassidis2005,Gebre-Egziabher2007,Springmann2012}, and the measurement error model of magnetometer calibration can be re-formulated into the form of Eq. (\ref{canonical2}). Another application in aircraft system identification is data compatibility analysis \cite{Klein2006}. Instrumentation errors from IMU and air data systems in both dynamic and measurement models can also be reformulated into this canonical form. Other applications such as attitude estimation \cite{Crassidis2007}, air data calibration \cite{Sun2019,Jurado2019} and stereo vision systems \cite{Chu2013,Johnson2017} also have similar models that can be re-formulated into this canonical form.    

In Appendix B, we show how two classical estimation problems can be reformulated into the canonical form shown in Eq. (\ref{canonical2}). The first example deals with the magnetometer calibration error model taken from Ref.  \cite{Gebre-Egziabher2007}. The second example deals with dynamic model equations for aircraft data compatibility analysis from Ref. \cite{Klein2006}. The first example only deals with a measurement error model assuming the time series is available. The second example considers unknown parameters from both dynamic and measurement error models.

It should be noted that though some problems can be recast into canonical form shown in Eq. (\ref{canonical2}), it does not mean the proposed method would necessarily be better than using conventional methods for parameter estimation. For example, even though the data compatibility problem can be solved by the proposed estimator, the proposed algorithm does not prove improve accuracy compared to the well-known  output-error method. What is unique about the proposed algorithm is that it may resolve the initial-value sensitivity problem if the measurements can be recast in suitable form, as demonstrated by the 5-hole Pitot tube calibration example in Sec. \ref{aeroEx}.

\section*{Appendix B: Application Examples} \label{AppB}

\setcounter{equation}{0}
\renewcommand{\theequation}{B\arabic{equation}}

\subsection{Magnetometer Calibration}

Consider the following magnetometer error equation \cite{Gebre-Egziabher2007}:

\begin{equation}
\begin{aligned}
{\bf h}^m &= {\bf C}_{\alpha} {\bf C}_{\eta} {\bf C}_{\lambda}{\bf h}^b + {\bf n} + {\bf v} \\
\end{aligned}
\end{equation}
where ${\bf C}_{\alpha}$, ${\bf C}_{\eta}$ and ${\bf C}_{\lambda}$ are soft-iron, misalignment and scale factor error matrices, respectively. $\renewcommand{\arraystretch}{0.7} {\bf h}^b =\small\left[\begin{array}{ccc} h^b_x & h^b_y & h^b_z\end{array}\small\right]^T$ is the true field magnetic vector in the body axes of the vehicle and $\boldsymbol{h}^m$ is the measured magnetic field vector. Null shifts or hard-iron biases are represented by the constant vector ${\bf n}$. The effect of wide-band, sampling, or sensor noise (uncorrelated noise) is represented by the vector ${\bf v}$. For details of this model, refer to Ref. \cite{Gebre-Egziabher2007}. Note that a more-complicated model can be found in Ref. \cite{Springmann2012}, where time-varying parameters are included in the measurement model. The objective is to estimate the following model parameters:

\begin{equation}
\begingroup 
\renewcommand{\arraystretch}{0.7}
\begin{aligned}
{\bs \xi}&= \begin{bmatrix}
\alpha_{ij} &
\eta_i &
\lambda_{i} &
n_{i}
\end{bmatrix}^T
\end{aligned} 
\endgroup
\end{equation}
where $i$ can be $x$, $y$, or $z$. 

With simple algebraic manipulation, the following canonical form can be obtained:

\begin{equation}
\begingroup 
\setlength\arraycolsep{1.5pt}
\renewcommand{\arraystretch}{0.7}
{\bf z} = {\bf h}^m  = \underbrace{\left[\begin{array}{c;{2pt/2pt}c}
	{\bf C}_{\alpha} {\bf C}_{\eta} \begin{pmatrix}
	h^b_x & & \\ & h^b_y &  \\ &  & h^b_z
	\end{pmatrix} & \boldsymbol{I}_{3 \times 3}
	\end{array}\right]}_{{\bf A}\left({\bs \xi}_2\right)} \underbrace{\left[\begin{array}{c}
	\lambda_x \\
	\lambda_y \\
	\lambda_z \\
	n_x \\
	n_y \\
	n_z \\
	\end{array}\right]}_{{\bs \xi}_{1}} + \underbrace{{\bf C}_{\alpha} {\bf C}_{\eta} \begin{bmatrix}
	h^b_x \\
	h^b_y \\
	h^b_z \\
	\end{bmatrix}}_{{\bf b}({\bs \xi}_2)}  + {\bf v}
\endgroup
\label{magnetometerModel}
\end{equation}
where $\setlength\arraycolsep{2pt}\renewcommand{\arraystretch}{0.7}  {\bs \xi} \mbox{\: are split into} \: {\bs \xi}_{1}=\small\left[\begin{array}{ccc} \lambda_{i} & 
n_{i} \end{array}\small\right]^T$ and $\setlength\arraycolsep{2pt}\renewcommand{\arraystretch}{0.7} {\bs \xi}_2=\small\left[\begin{array}{ccc} \alpha_{ij} &
\eta_i \end{array}\small\right]^T$. It can be clearly seen that Eq. (\ref{magnetometerModel}) has the same form as Eq. (\ref{canonical2}).

\subsection{Aircraft Data Compatibility}\label{airdataExample}

Another common application in aerospace engineering is data compatibility analysis. In particular, aircraft data compatibility analysis is a process of estimating and removing systematic instrumentation errors that create kinematic inconsistencies in the measured sensor data. The classic example from Ref. \cite{Klein2006} is used to show how this application can also be transformed into the canonical form. The typical states ${\bf x}$, input ${\bf u}$, measurement ${\bf z}$ and set of typical parameters ${\bs \xi}$ for this problem are given by the following:
\begin{subequations}
	\begingroup 
	\setlength\arraycolsep{1.5pt}
	\begin{align}
	{\bf x} &= \begin{bmatrix} u & v & w & \phi & \theta & \psi \end{bmatrix}^T \\
	{\bf u} &= \begin{bmatrix} a_x & a_y & a_z & p & q & r \end{bmatrix}^T \\
	{\bf z} &= \begin{bmatrix} V_a & \beta & \alpha & \phi & \theta & \psi \end{bmatrix}^T\\ 
	{\bs \xi} &= \begin{bmatrix}
	b_{a_x} &
	b_{a_y}&
	b_{a_z} &
	b_{p} &
	b_{q} &
	b_{r} &
	\lambda_{V_a} &
	\lambda_{\alpha} &
	\lambda_{\beta} &
	b_{V_a} &
	b_{\alpha} &
	b_{\beta} &
	\lambda_{\phi} &
	\lambda_{\theta} &
	\lambda_{\psi} &
	b_{\phi} &
	b_{\theta} &
	b_{\psi} &
	\end{bmatrix}^T
	\end{align}
	\endgroup
\end{subequations}
where $\lambda_{\phi}, \lambda_{\theta},\lambda_{\psi}, b_{\phi}, b_{\theta}, b_{\psi}$ are scale factors and biases of Euler angles in addition to the parameters introduced in the earlier sections.  
The dynamic model for data compatibility analysis are:
\begin{equation}
\begingroup 
\setlength\arraycolsep{1.2pt}
\renewcommand{\arraystretch}{0.7}
\begin{aligned}
\begin{bmatrix}
\dot{u} \\
\dot{v} \\
\dot{w} \\
\end{bmatrix} &= \begin{bmatrix}
0 & r + b_{r} & -(q+b_q) \\
-(r+ b_r) & 0 & p+b_p \\
q+b_q & -(p+b_p) & 0 
\end{bmatrix}\begin{bmatrix}
u \\
v \\
w \\
\end{bmatrix} \\
&+ \begin{bmatrix}
-g\sin\theta + a_x + b_{a_x} \\
g\sin\phi \cos\theta +a_y + b_{a_y} \\
g\cos\phi \cos\theta +a_z + b_{a_z} \\
\end{bmatrix} \\
\begin{bmatrix}
\dot{\phi} \\
\dot{\theta} \\
\dot{\psi} \\
\end{bmatrix} &= \begin{bmatrix}
1 & \sin\phi \tan\theta & \cos\phi \tan\theta \\
0 & \cos \phi & -\sin \phi \\
0 & \dfrac{\sin \phi}{\cos \theta}  & \dfrac{\cos \phi}{\cos \theta}  \\
\end{bmatrix}
\begin{bmatrix}
p + b_p \\
q + b_q \\
r + b_r \\
\end{bmatrix}
\end{aligned}
\endgroup
\label{kinematic}
\end{equation}
Finally, the measurement model outputs ${\bf z}$ are the airspeed, air flow angles, and Euler angles:
\begin{equation}
\begingroup 
\setlength\arraycolsep{1.0pt}
\renewcommand{\arraystretch}{0.7}
\begin{aligned}
V_a &= (1+\lambda_V)\sqrt{u^2 +v^2+w^2} + b_{V_a} + v_{V_a}\\
\beta &= (1+\lambda_{\beta}) \sin^{-1}(v / \sqrt{u^2 +v^2+w^2}) + b_{\beta} + v_{\beta} \\
\alpha &= (1+\lambda_{\alpha}) \tan^{-1}(w / u) + b_{\alpha} + v_{\alpha} \\
\phi &= (1+\lambda_{\phi})\phi + b_{\phi} + v_{\phi} \\
\theta &= (1+\lambda_{\theta})\theta + b_{\theta} + v_{\theta} \\
\psi &= (1+\lambda_{\psi})\psi + b_{\psi} + v_{\psi} \\
\end{aligned}
\endgroup
\end{equation}
With some algebraic manipulation, the measurement output model can be recast into the canonical form as follows:
\begin{equation}
\begingroup 
\setlength\arraycolsep{0.7pt}
\renewcommand{\arraystretch}{0.7}
\begin{aligned}
\boldsymbol{z}_k = \begin{bmatrix}
V_a \\
\beta \\
\alpha \\
\phi \\
\theta \\
\psi \\
\end{bmatrix}_k  &= \underbrace{
	\left[\begin{array}{cc;{2pt/2pt}cc;{2pt/2pt}cc;{2pt/2pt}cc;{2pt/2pt}cc;{2pt/2pt}cc}	
	V_{a_{k}} & 1 & & & & & & & & & & 	\\
	&  & \beta_{k}& 1 & & & & & & & & \\
	& & & & \alpha_{k} & 1 & & & & &  &\\   	
	& & & &   & & \phi_{k} & 1 & & & & \\   
	& & & &   & &  &  & \theta_{k} & 1 & \\   
	& & & &   & &  &  &  &  & \psi_{k} & 1 \\  
	\end{array}\right]}_{{\bf A}({\bf x}, {\bf u},{\bs \xi}_2)} \underbrace{\begin{bmatrix}
	\lambda_{V_a} \\
	b_{V_a} \\
	\lambda_{\beta} \\
	b_{\beta} \\
	\lambda_{\alpha} \\
	b_{\alpha} \\
	\lambda_{\phi} \\
	b_{\phi} \\
	\lambda_{\theta} \\
	b_{\theta} \\	
	\lambda_{\psi} \\
	b_{\psi} \\
	\end{bmatrix}}_{{\bs \xi}_{1}} + \underbrace{\begin{bmatrix}
	V_{k}   \\
	\beta_{k}  \\
	\alpha_{k}   \\
	\phi_{k} \\
	\theta_{k} \\
	\psi_{k} \\ 
	\end{bmatrix}}_{{\bf b}({\bf x}, {\bf u},{\bs \xi}_2)}  + \underbrace{\begin{bmatrix}
	v_{V_a} \\
	v_{\beta} \\
	v_{\alpha} \\
	v_{\phi} \\
	v_{\theta} \\
	v_{\psi} \\
	\end{bmatrix}_k}_{{\bf v}_k}
\end{aligned}
\endgroup
\label{airdataModel}
\end{equation}
where ${\bs \xi}$ are split into  ${\bs \xi}_{1}$ and ${\bs \xi}_2$ shown as follows: 

\begin{subequations}
	\begingroup 
	\setlength\arraycolsep{1.5pt}
	\renewcommand{\arraystretch}{0.7}
	\begin{align}
	{\bs \xi}_{1} &= \begin{bmatrix}
	\lambda_{V} &
	\lambda_{\alpha} &
	\lambda_{\beta} &
	b_{V} &
	b_{\alpha} &
	b_{\beta} &
	\lambda_{\phi} &
	\lambda_{\theta} &
	\lambda_{\psi} &
	b_{\phi} &
	b_{\theta} &
	b_{\psi} &
	\end{bmatrix}^T \\
	{\bs \xi}_2 &= \begin{bmatrix}
	b_{a_x} &
	b_{a_y}&
	b_{a_z} &
	b_{p} &
	b_{q} &
	b_{r} &
	\end{bmatrix}^T
	\end{align}
	\endgroup
\end{subequations}

The airspeed $V_k$, angle-of-sideslip $\beta_k$ and angle-of-attack $\alpha_k$ in ${\bf A}({\bf x},{\bf u},{\bs \xi}_2)$ and ${\bf b}({\bf x}, {\bf u},{\bs \xi}_2)$ are calculated by the state ${\bf x}_k$ shown in Eq. (\ref{airDataMesurement}).
\begin{equation}
\begin{aligned}
V_{k} &= \sqrt{u_{k}^2 +v_{k}^2+w_{k}^2} \\
\beta_{k} &= \sin^{-1}(v_{k} / \sqrt{u_{k}^2+v_{k}^2+w_{k}^2}) \\
\alpha_{k} &=  	\tan^{-1}(w_{k} / u_{k}) \\	   
\end{aligned}
\label{airDataMesurement}
\end{equation}

In order to use the proposed estimator, all the states ${\bf x}_k$ for $k = 1,...,N$ have to be known, which is a downside of this algorithm. Also, data compatibility problems are not particularly sensitive to initial conditions. It was well-known that the zero initial condition is sufficient to solve such problems via output-error. Nevertheless, the proposed estimator is a viable and convenient alternative.   

Both magnetometer calibration and aircraft data compatibility analysis examples reveal that a common nonlinear parameter estimation problem can be transformed into an affine linear model as shown in Eq. (\ref{canonical3}). The unknown parameters are separated into two sets with simple algebraic manipulation. With this canonical form, the proposed estimator can be used to solve the parameter estimation problem accurately and consistently.

\bibliography{mybib}

\begin{thebibliography}{28}
\newcommand{\enquote}[1]{``#1''}
\providecommand{\natexlab}[1]{#1}
\providecommand{\url}[1]{\texttt{#1}}
\providecommand{\urlprefix}{URL }
\expandafter\ifx\csname urlstyle\endcsname\relax
  \providecommand{\doi}[1]{doi:\discretionary{}{}{}#1}\else
  \providecommand{\doi}{doi:\discretionary{}{}{}\begingroup
  \urlstyle{rm}\Url}\fi

\bibitem[{Stengel(1994)}]{Stengel}
Stengel, R.~F., \emph{Optimal Control and Estimation}, Dover, Mineola, NY,
  1994.
\newblock Chatper 2,4.

\bibitem[{Simon(2006)}]{Simon2006}
Simon, D., \emph{Optimal State Estimation: Kalman, H Infinity, and Nonlinear
  Approaches}, Wiley-Interscience, Hoboken, N.J., USA, 2006.
\newblock Chatper 5-14.

\bibitem[{Kay(1993)}]{Kay1993}
Kay, S.~M., \emph{Fundamentals of Statistical Signal Processing: Estimation
  Theory}, Prentice-Hall, Inc., Upper Saddle River, NJ, USA, 1993.
\newblock Chatper 7.

\bibitem[{Jategaonkar(2006)}]{Jategaonkar2006}
Jategaonkar, R., \emph{Flight Vehicle System Identification: A Time Domain
  Methodology}, Vol. 216, AIAA, 2006.
\newblock Chatper 4.

\bibitem[{Klein and Morelli(2006)}]{Klein2006}
Klein, V., and Morelli, E., \emph{Aircraft System Identification: Theory and
  Practice}, AIAA education series, American Institute of Aeronautics and
  Astronautics, 2006.
\newblock Chatper 6.

\bibitem[{Grauer(2015)}]{Grauer2015}
Grauer, J.~A., \enquote{Real-Time Data-Compatibility Analysis Using
  Output-Error Parameter Estimation,} \emph{Journal of Aircraft}, Vol.~52,
  No.~3, 2015, pp. 940--947.
\newblock \doi{10.2514/1.c033182}.

\bibitem[{Groves(2013)}]{Groves}
Groves, P.~D., \emph{Principles of GNSS, Inertial, and Multisensor Integrated
  Navigation Systems}, 2\textsuperscript{nd} ed., Artech House, Boston, MA,
  USA, 2013.
\newblock Chatper 4.

\bibitem[{Lawton and Sylvestre(1971)}]{Lawton1971}
Lawton, W.~H., and Sylvestre, E.~A., \enquote{Elimination of Linear Parameters
  in Nonlinear Regression,} \emph{Technometrics}, Vol.~13, No.~3, 1971, pp.
  461--467.
\newblock \doi{10.2307/1267160}.

\bibitem[{Guttman et~al.(1973)Guttman, Pereyra, and Scolnik}]{Guttman1973}
Guttman, I., Pereyra, V., and Scolnik, H.~D., \enquote{Least Squares Estimation
  for a Class of Non-Linear Models,} \emph{Technometrics}, Vol.~15, No.~2,
  1973, pp. 209--218.
\newblock \doi{10.2307/1266982}.

\bibitem[{Golub and Pereyra(1973)}]{Golub1973}
Golub, G., and Pereyra, V., \enquote{The Differentiation of Pseudo-Inverses and
  Nonlinear Least Squares Problems Whose Variables Separate,} \emph{SIAM J.
  Numer. Anal.}, Vol.~10, No.~2, 1973, pp. 413--432.
\newblock \doi{10.1137/0710036}.

\bibitem[{Bates and Watts(1988)}]{Bates1988}
Bates, D., and Watts, D., \emph{Nonlinear Regression Analysis and Its
  Applications}, Wiley Series in Probability and Statistics - Applied
  Probability and Statistics Section Series, Wiley, 1988.
\newblock \MakeLowercase{p}p. 78-79.

\bibitem[{Haupt et~al.(1996)Haupt, Kasdin, Keiser, and Parkinson}]{Haupt1996}
Haupt, G.~T., Kasdin, N.~J., Keiser, G.~M., and Parkinson, B.~W.,
  \enquote{Optimal recursive iterative algorithm for discrete nonlinear
  least-squares estimation,} \emph{Journal of Guidance, Control, and Dynamics},
  Vol.~19, No.~3, 1996, pp. 643--649.
\newblock \doi{10.2514/3.21669}.

\bibitem[{Alonso and Shuster(2002)}]{Alonso2002}
Alonso, R., and Shuster, M.~D., \enquote{TWOSTEP: A fast robust algorithm for
  attitude-independent magnetometer-bias determination,} \emph{The Journal of
  the Astronautical Sciences}, Vol.~50, 2002, pp. 433--451.

\bibitem[{Hauer et~al.(1990)Hauer, Demeure, and Scharf}]{Hauer1990}
Hauer, J.~F., Demeure, C.~J., and Scharf, L.~L., \enquote{Initial results in
  Prony analysis of power system response signals,} \emph{IEEE Transactions on
  Power Systems}, Vol.~5, No.~1, 1990, pp. 80--89.
\newblock \doi{10.1109/59.49090}.

\bibitem[{Marquardt(1963)}]{Marquardt1963}
Marquardt, D., \enquote{An Algorithm for Least-Squares Estimation of Nonlinear
  Parameters,} \emph{Journal of the Society for Industrial and Applied
  Mathematics}, Vol.~11, No.~2, 1963, pp. 431--441.
\newblock \doi{10.1137/0111030}.

\bibitem[{Crassidis et~al.(2005)Crassidis, Lai, and Harman}]{Crassidis2005}
Crassidis, J.~L., Lai, K.-L., and Harman, R.~R., \enquote{Real-Time
  Attitude-Independent Three-Axis Magnetometer Calibration,} \emph{Journal of
  Guidance, Control, and Dynamics}, Vol.~28, No.~1, 2005, pp. 115--120.
\newblock \doi{10.2514/1.6278}.

\bibitem[{MathWorks(2019)}]{fmincon}
MathWorks, \enquote{fmincon,}
  https://www.mathworks.com/help/optim/ug/fmincon.html, 2019.
\newblock Accessed: 2019-10-05.

\bibitem[{Sun and Gebre-Egziabher(2019)}]{Sun2019}
Sun, C.~D., Kerry.~Regan, and Gebre-Egziabher, D., \enquote{A GNSS/IMU-Based
  5-Hole Pitot Tube Calibration Algorithm,} \emph{AIAA Scitech 2019 Forum},
  AIAA SciTech Forum, AIAA, 2019.
\newblock \doi{10.2514/6.2019-0360}.

\bibitem[{Owens et~al.(2006)Owens, Cox, and Morelli}]{Owens2006}
Owens, B., Cox, D., and Morelli, E., \enquote{Development of a Low-Cost
  Sub-Scale Aircraft for Flight Research: The FASER Project,} \emph{American
  Institute of Aeronautics and Astronautics}, 2006.
\newblock \doi{10.2514/6.2006-3306}.

\bibitem[{Ros(1988)}]{Rosemount1988}
\emph{Rosemount Model 858 Flow Angle Sensors}, Bulletin 1014, Rosemount Inc.,
  Burnsville, MN, 1988.

\bibitem[{umn(2019)}]{umnUAVlab}
\enquote{Unversity of Minnesota UAV Laboratories,}
  https://www.uav.aem.umn.edu/resources/goldy-iii, 2019.
\newblock Accessed: 2019-10-01.

\bibitem[{Sun et~al.(2019)Sun, Regan, and Gebre-Egziabher}]{Sun2019a}
Sun, K., Regan, C.~D., and Gebre-Egziabher, D., \enquote{Observability and
  Performance Analysis of a Model-Free Synthetic Air Data Estimator,}
  \emph{Journal of Aircraft}, Vol.~56, No.~4, 2019, pp. 1471--1486.
\newblock \doi{10.2514/1.c035290}.

\bibitem[{Gebre-Egziabher(2007)}]{Gebre-Egziabher2007}
Gebre-Egziabher, D., \enquote{Magnetometer Autocalibration Leveraging
  Measurement Locus Constraints,} \emph{Journal of Aircraft}, Vol.~44, No.~4,
  2007, pp. 1361--1368.
\newblock \doi{10.2514/1.27118}.

\bibitem[{Springmann and Cutler(2012)}]{Springmann2012}
Springmann, J.~C., and Cutler, J.~W., \enquote{Attitude-Independent
  Magnetometer Calibration with Time-Varying Bias,} \emph{Journal of Guidance,
  Control, and Dynamics}, Vol.~35, No.~4, 2012, pp. 1080--1088.
\newblock \doi{10.2514/1.56726}.

\bibitem[{Crassidis et~al.(2007)Crassidis, Markley, and Cheng}]{Crassidis2007}
Crassidis, J.~L., Markley, F.~L., and Cheng, Y., \enquote{Survey of Nonlinear
  Attitude Estimation Methods,} \emph{Journal of Guidance, Control, and
  Dynamics}, Vol.~30, No.~1, 2007, pp. 12--28.
\newblock \doi{10.2514/1.22452}.

\bibitem[{Jurado and McGehee(2019)}]{Jurado2019}
Jurado, J.~D., and McGehee, C.~C., \enquote{Complete Online Algorithm for Air
  Data System Calibration,} \emph{Journal of Aircraft}, Vol.~56, No.~2, 2019,
  pp. 517--528.
\newblock \doi{10.2514/1.c034964}.

\bibitem[{Chu et~al.(2013)Chu, F.~Adhika~Pradipta, and
  Gebre-Egziabher}]{Chu2013}
Chu, C., F.~Adhika~Pradipta, L., and Gebre-Egziabher, D., \enquote{Dual
  Hypothesis Filter for Robust INS/Camera Fusion,} \emph{Proceedings of the
  2013 International Technical Meeting of The Institute of Navigation}, San
  Diego, CA, USA, 2013, pp. 792--802.
\newblock \doi{10.2514/6.2019-0360}.

\bibitem[{Johnson et~al.(2017)Johnson, Nykl, and Raquet}]{Johnson2017}
Johnson, D.~T., Nykl, S.~L., and Raquet, J.~F., \enquote{Combining Stereo
  Vision and Inertial Navigation for Automated Aerial Refueling,} \emph{Journal
  of Guidance, Control, and Dynamics}, Vol.~40, No.~9, 2017, pp. 2250--2259.
\newblock \doi{10.2514/1.g002648}.

\end{thebibliography}

\end{document}